\documentclass[longauth]{aa}
\usepackage{lineno}
\usepackage{longtable}
\usepackage{lscape}
\usepackage{booktabs}
\usepackage{multirow}
\usepackage{natbib}
\bibpunct{(}{)}{;}{a}{}{,}
\usepackage{pifont}

\usepackage{orcidlink}
\usepackage{xcolor}
\usepackage{graphicx}
\usepackage{txfonts}
\usepackage{hyperref}
\usepackage[normalem]{ulem}
\usepackage{subcaption}
\usepackage{threeparttablex}
\usepackage{longtable}
\usepackage{placeins}
\usepackage{float}

\begin{document}

\title{The (missing) gamma-ray counterparts to EP events}

   \subtitle{Part I: Search for keV-GeV 
   counterparts of \\ Einstein Probe Fast X-ray Transients}

\author{{M. E.} Ravasio\orcidlink{0000-0003-3193-4714}\inst{1,2,3,4},
          {A.} Chopra\orcidlink{0000-0001-5679-0695}\inst{5},
          {P. G.} Jonker\orcidlink{0000-0001-5679-0695}\inst{3},
          {Q.} Wu\orcidlink{0009-0002-9275-715X}\inst{7},
          {E.} Burns\orcidlink{0000-0002-2942-3379}\inst{8},
          {B.} Banerjee\orcidlink{0000-0002-8008-2485}\inst{5,6},
          {G.} Oganesyan\orcidlink{0000-0001-9765-1552}\inst{5,6},
          {S.} Ascenzi\orcidlink{0000-0001-5116-6789}\inst{5,6},
          {J. T.} Palmerio\orcidlink{0000-0002-9408-1563}\inst{9},
          {F.} Daigne\inst{10},
          {J.} Wood\orcidlink{0000-0001-9012-2463}\inst{11},
          {R.} Hamburg\orcidlink{0000-0003-0761-6388}\inst{12},
          {A. J.} Levan\orcidlink{0000-0001-7821-9369}\inst{3,13},
          {J. A.} Quirola Vasquez\orcidlink{0000-0001-8602-4641}\inst{3},
          {J.} S\'anchez-Sierras\orcidlink{0000-0003-2276-4231}\inst{3},
          {J. N. D.} van Dalen\orcidlink{0009-0007-6927-7496}\inst{3},
          {A. P. C.}  van Hoof\orcidlink{0009-0005-5404-2745}\inst{3},
          {H.} Sun\orcidlink{0000-0002-9615-1481}\inst{7},
          {W.} Yuan\inst{7},
          {N.} Sarin\orcidlink{0000-0003-2700-1030}\inst{14,15},
          {A.} Martin-Carrillo\orcidlink{0000-0001-5108-0627}\inst{16}, 
          {D.} Adrien\orcidlink{0009-0002-9333-0182}\inst{9},
          {C.} Cai\orcidlink{0000-0002-6540-2372}\inst{17},
          {Y.} Zhang\orcidlink{0000-0001-5348-7033}\inst{18},
          {S.} Ronchini\orcidlink{0000-0003-0020-687X}\inst{5,6}, 
          {M.} Branchesi\orcidlink{0000-0003-1643-0526}\inst{5,6}}

   \institute{Institute of Space Sciences (ICE, CSIC), Campus UAB, Carrer de Can Magrans s/n, Barcelona, E-08193, Spain
        \and 
        Institut d’Estudis Espacials de Catalunya (IEEC), Edifici RDIT, Campus UPC, Castelldefels (Barcelona), E-08860, Spain
        \and
        Department of Astrophysics/IMAPP, Radboud University, 6525 AJ Nijmegen, The Netherlands 
        \and 
        INAF -- Osservatorio Astronomico di Brera, via Emilio Bianchi 46, I-23807 Merate (LC), Italy
        \and
        Gran Sasso Science Institute, viale F. Crispi 7, L'Aquila, I-67100, Italy
        \and
        INFN -- Laboratori Nazionali del Gran Sasso, L'Aquila, I-67100, Italy 
        \and
        National Astronomical Observatories, Chinese Academy of Sciences, Datun Road A20, Beijing 100012, China 
        \and
        Department of Physics \& Astronomy, Louisiana State University, Baton Rouge, LA 70803, USA 
        \and
        CEA Paris-Saclay, IRFU/DAp-AIM, 91191 Gif sur Yvette, France 
        \and
        Sorbonne Universit\'e, CNRS, UMR 7095, Institut d’Astrophysique de Paris, 98 Bis bd Arago, 75014, Paris, France 
        \and
        ST12 Astrophysics Branch, NASA Marshall Space Flight Center, Huntsville, AL 35812, USA
        \and
        Science and Technology Institute, Universities Space Research Association, Huntsville, AL 35805, USA
        \and
        Department of Physics, University of Warwick, Coventry, CV4 7AL, UK 
        \and
        Kavli Institute for Cosmology, University of Cambridge, Madingley Road, Cambridge, CB3 0HA, UK
        \and
        Institute of Astronomy, University of Cambridge, Madingley Road, Cambridge, CB3 0HA, UK
        \and
        School of Physics and Centre for Space Research, University College Dublin, Belfield, Dublin, Ireland 
        \and
        College of Physics and Hebei Key Laboratory of Photophysics Research and Application, Hebei Normal University, Shijiazhuang, Hebei 050024, People’s Republic of China
        \and
        School of Physics and Electronic Science, Guizhou Normal University, Guiyang 550001, People's Republic of China
   }

\authorrunning{M.~E.\ Ravasio et al.}

   \date{Received ; accepted }

 
  \abstract{}
{While the connection of some Fast X-ray Transients (FXTs) detected by \textit{Einstein Probe} (EP) with Gamma-ray Bursts (GRBs) is increasingly compelling, most lack $\gamma$-ray counterparts, leaving open the question of whether and how they are related to the GRB phenomenon.} 
{We performed a systematic broadband search for $\gamma$-ray counterparts to the 72 transients detected by EP during its first year of operations, using dedicated targeted searches in \textit{Fermi}/GBM and \textit{Fermi}/LAT data, complemented by information from SVOM/GRM, SVOM/ECLAIRs, Swift/BAT and Konus-Wind.
For the visible events without a $\gamma$-ray counterpart, we derived simultaneous $3\sigma$ flux upper limits, during the full duration of the event and its peak time, in the 10--1000\,keV and the 0.1--1\, GeV band.
} 
{We find that 23 events (32\%) have a $\gamma$-ray counterpart (seven sub-threshold), while 49 events (68\%) have no detected $\gamma$-ray counterpart, despite most being visible.
For events with a $\gamma$-ray counterpart, when the X-ray flux is measured over the $\gamma$-ray $T_{\rm 90}$, it is systematically higher than the flux measured over the full X-ray duration, with 
flux ratios $1 \lesssim F_{\rm [10-1000\,keV]}/F_{\rm [0.5-4\,keV]} \lesssim 100$, suggesting that $\gamma$-rays are preferentially detected during the brightest X-ray phases.
For events without $\gamma$-ray counterpart, the most stringent \textit{Fermi}/GBM upper limits imply spectral peak energies $E_{\rm peak} \lesssim 30$--100\,keV and, for the 14 events with measured redshifts, isotropic-equivalent luminosities $L_{\rm iso}\lesssim1\times10^{48}-9\times10^{51}$ erg s$^{-1}$.  
Assuming the same correlations established for classical GRBs holds in this regime, these lower energetics imply lower bulk Lorentz factors ($\Gamma_0 \lesssim 100$) for EP FXTs compared to those of classical GRBs, as independently confirmed in a few cases by afterglow constraints.
No significant GeV emission was detected by \textit{Fermi}/LAT for any visible EP transient, yielding $1 \lesssim F_{0.1\text{-}1\,{\rm GeV}}/F_{0.5\text{-}4\,{\rm keV}} \lesssim 10$, which places the first observational constraints on a potential synchrotron self-Compton component for this new population of X-ray transients.}
{While the origin of individual EP transients without $\gamma$-ray counterparts remains uncertain, these high-energy constraints suggest that many of them could be compatible with the GRB population, extending it toward lower $E_{\rm peak}$, lower luminosities, and lower bulk Lorentz factors.
\textit{Einstein Probe} is finally providing the opportunity to systematically characterize these unexplored regimes of jet physics.}


   \maketitle


\section{Introduction}

For more than five decades, gamma-ray bursts (GRBs) have been primarily discovered and characterized through their prompt $\gamma$-ray  emission. From their discovery by the \textit{Vela} satellites \citep{1973Klebesadel}, decades of observations with dedicated $\gamma$-ray instruments, from BATSE to \textit{Swift}/BAT and \textit{Fermi}/GBM, have built a rich picture of the GRB phenomenology in the hard X-ray and $\gamma$-ray bands, establishing e.g. the bimodal duration distribution \citep{1993Kouveliotou}, the discovery of their cosmological origin (thanks to rapid X-ray localization,  \citealt{1997Costa,1997VanParadijs}), the spectral correlations \citep{2004ApJ...609..935Y, 2002Amati}, and the connection to their progenitors: the collapse of massive stars for long GRBs \citep[e.g.][]{2001Woosley, 1998Galama}, and the merger of compact objects for short GRBs \citep[e.g.][]{2017Abbott, 2017Granot, 2021Ascenzi}.

In contrast, the X-ray band during the prompt phase has remained comparatively underexplored, despite its importance for probing GRB physics, e.g. the central engine activity, the radiative mechanism \citep[e.g.][]{Oganesyan2018}, and the jet structure, among others. The limited field of view and low-energy sensitivity of past X-ray missions have hindered the systematic exploration of the X-ray transient sky.
Among the softest members of the GRB population known to date, X-ray flashes \citep[XRFs;][]{Heise2001, Frontera2004}, 
events whose prompt emission peaks below $\sim$30\,keV, were first identified by BeppoSAX, but have remained under-sampled ever since. Their physical origin is still debated, with proposed scenarios including on-axis weak jets \citep[e.g.][]{Lamb2004}, off-axis relativistic jets \citep[e.g.][]{Yamazaki2002}, and relativistic shock breakouts \citep[e.g.][]{Wang2007}.
While \textit{Swift}/XRT has revolutionised in many ways our understanding of GRBs, including the study of XRF \citep{Sakamoto2008} and the sampling of the transition between the prompt and early afterglow phases, simultaneous X-ray observations of the beginning of the prompt emission are still rare. 
As a result, the softest tail of the GRB distribution remains largely inaccessible to systematic studies.

In parallel, several fast X-ray transients (FXTs) have been discovered over the past two decades through archival searches in X-ray catalogs of \textit{Chandra} and \textit{XMM-Newton} \citep{2008Soderberg, 2013Jonker, 2017Bauer, 2023QuirolaVasquez, AlpLarsson2020}. These searches revealed a class of long-lived ($\sim$ few to tens of ks) X-ray flares with no obvious $\gamma$-ray counterparts. Their durations, fluxes, spectra, and temporal properties suggest a connection with relativistic explosions; however, the lack of counterparts at other wavelengths and robust redshift estimates in most cases (despite thorough searches for host galaxies, see \citealt{2021Jonker, 2022Eappachen, 2023Eappachen, Quirola-Vasquez2026}) has prevented a clear determination of their physical nature. A wide range of progenitor models has therefore been tested, including supernova shock breakouts \citep{AlpLarsson2020}, compact binary mergers producing long-lived magnetars \citep{2019Sun,2024QuirolaVasquez, 2019Xue}, off-axis GRBs afterglows \citep{2021Sarin,Wichern2024}, and tidal disruption events, but their nature remains largely unconstrained. Yet, these discoveries hinted at the existence of a rich and unexplored population of X-ray transients, potentially linked to phenomena traditionally studied in $\gamma$-rays.

This observational landscape has changed significantly with the advent of the \textit{Einstein Probe} mission (EP; \citealt{Yuan2022}) and, soon after, \textit{SVOM} \citep{Wei2023}. Thanks to their sensitivity at soft X-ray energies, the Wide-field X-ray Telescope (WXT; 0.5--4 keV) on board EP and the ECLAIRs instrument (4--150 keV) on board SVOM have opened a new window on the poorly explored soft high-energy transient sky. In particular, EP/WXT, employing novel lobster-eye optics, combines a large field of view ($\sim$3600 deg$^2$) with a sensitivity of $\sim2.6\times10^{-11}$ erg cm$^{-2}$ s$^{-1}$ in a 1 ks exposure \citep{Yuan2022}, allowing the discovery and rapid localisation of faint X-ray transients. 
Several of the EP FXTs have been directly associated with GRBs through contemporaneous detections by $\gamma$-ray satellites \citep[e.g.][]{Yin2024, Jiang2025}, even at redshift $\sim5$ \citep{Levan2024, Liu2025}, while others showed no $\gamma$-ray counterparts \citep[e.g.][]{Ravasio2024GCN.36725....1R,Ravasio2025GCN.41273....1R,Ravasio2025GCN.41815....1R,Zhang2025_upperlimits}. Interestingly, despite the lack of $\gamma$-ray detections in many cases, EP transients often exhibit late-time lightcurve properties consistent with those of GRB afterglows, suggesting a potential physical connection even when the prompt gamma-ray emission is not observed \citep{Sun2025_EP240414a, Quirola-Vasquez2026, Cotter2026, Gianfagna2025}. Indeed, among events without a $\gamma$-ray counterpart, many have been linked to the collapsar channel through the emergence of broad-lined Type Ic supernovae (Ic-BL SNe), similar to those accompanying nearby long GRBs \citep[e.g.][]{vanDalen2025, Rastinejad2025, Li2025, Quirola-Vasquez2026, vanHoof2026}.  The inferred formation rate and luminosity function of FXTs are reported to be consistent with a massive-star progenitor channel \citep{Guo2025}, while \citealt{OConnor2025} found that their redshift distribution is statistically consistent with that of long GRBs.
Evidence for an association with compact object mergers remains more elusive, with the exception of EP250207b \citep{Jonker2026} (later found to be at z=3.2, van Hoof et al., in prep), and 
EP250704a, which was coincident with a short GRB \citep{Fraija2026, Li2026}. 
Other proposed physical origins for EP FXTs include  TDEs, as suggested for GRB 250702B/EP250702a \citep{Neights2026, Gompertz2026, Levan2025, Oganesyan2025, Granot2026, Li2026_BDE}, and supernova shock breakout (SBO), as in the case of EP260321a \citep{Yuan2026_SBO,Martin-Carrillo2026_SBO,Rastinejad2026, Chen2026, OCONNOR2026_SBO}.
The physical nature of a large fraction of EP FXTs without a $\gamma$-ray counterpart remains, however, fundamentally unconstrained.

Given the emerging evidence for a connection between FXTs and the GRB population, it is interesting to systematically search for and characterise prompt $\gamma$-ray counterparts to EP-detected events, to investigate their association and assess whether FXTs represent an extension of classical GRBs or a distinct class of relativistic transients.
Starting from the first EP catalog of extragalactic transients (Q. Wu et al., in prep), in this paper we present a systematic search for $\gamma$-ray counterparts to the extragalactic transients detected by EP/WXT during its first year of operation (February 2024 -- February 2025), using contemporaneous observations from \textit{Fermi}/GBM and \textit{Fermi}/LAT.  
The paper is structured as follows. Section~\ref{sec:sampleselection} describes the sample selection and the search methods for \textit{Fermi}/GBM and \textit{Fermi}/LAT. The results of the $\gamma$-ray search are presented in Section~\ref{sec:results}, including the multi-wavelength flux comparisons and 
the derived upper limits. Section~\ref{sec:discussion} discusses the connection to the broader GRB population, the implications for the spectral properties, energetics, jet dynamics and the constraints on potential GeV spectral components. We summarise our main conclusions in Section~\ref{sec:conclusions}. Throughout this work, we assume $\Lambda$CDM cosmology with $H_0 = 67.4$\,km\,s$^{-1}$\,Mpc$^{-1}$ and $\Omega_{\rm m} = 0.315$ \citep{Planck2020}.

\section{Methods}\label{sec:sampleselection}
\subsection{Sample selection}

Our sample selection started from the complete set of 97 publicly announced transient events detected by the Einstein Probe Wide-field X-ray Telescope (EP/WXT) during its first year of operations, spanning from February 2024 to February 2025. As our work focuses on the newly discovered extragalactic transients, from the initial list of 97 sources we excluded stellar flares, unconfirmed detections, and previously known sources such as X-ray binaries and blazars. The final sample comprises 72 transients detected by EP/WXT in the first year of activity, as reported in the first catalogue of EP extragalactic transients (Q. Wu et al., in prep). 
From this EP first-year catalogue, we collected the X-ray trigger time ($T_{0, \rm X\text{-}ray}$), the total X-ray duration ($T_{\rm 100,X\text{-}ray}$)\footnote{The total X-ray duration is defined as the time interval between the first and the last Bayesian block with SNR > 3 in the background-corrected light curve (see Q. Wu et al. 2026, in prep., for details).}, the time-averaged unabsorbed flux with its $1\sigma$ uncertainties in the 0.5--4\,keV energy band ($F_{\rm avg,X\text{-}ray}$), the time of the X-ray light curve peak $T_{\rm peak,X\text{-}ray}$, and the unabsorbed flux at peak time with its $1\sigma$ uncertainties ($F_{\rm peak,X\text{-}ray}$).
For this sample, we performed a targeted search for temporally and spatially\footnote{For GBM-detected events, we computed $p_{\rm spatial}$, the probability that the GBM localization is genuinely associated with the WXT position (as opposed to a chance coincidence) following \citet{Fletcher2024}, via the GDT-Fermi \texttt{source\_probability} function, finding it to have an average value of 96.3\%.} coincident $\gamma$-ray counterparts using data from both instruments aboard \textit{Fermi}: the Gamma-ray Burst Monitor \citep[GBM;][]{Meegan2009_GBM} and the Large Area Telescope \citep[LAT;][]{Atwood2009_LAT}. The criteria adopted for the selection of the corresponding subsamples are described in the following sections.

In addition to the \textit{Fermi}-based analysis, for those events not visible in Fermi, we complemented our search with information from other missions. For example, a targeted search of the Konus-Wind data (the details of which will be presented in Wu et al. 2026, in preparation) led to the identification of two additional tentative $\gamma$-ray counterparts (see Sec.\ref{sec:GBM_results}), for the events EP240202a and EP240703b. We report this tentative association in this work and include them in the GRB category, but the lack of detailed information on the search method, together with the unavailability of the corresponding $\gamma$-ray properties (e.g. flux, $E_{\rm peak}$, $T_{90}$), leads us to exclude these two events from the comparative analysis presented in the following sections (e.g. flux distributions and associated statistical tests). We note that neither event has a measured redshift, so their exclusion does not affect the $L_{\rm iso}$ and $E_{\rm iso}$ estimates discussed later in this work.
Moreover, we also considered the results of \citet{Zhang2025_upperlimits}, who performed a similar $\gamma$-ray search using GECAM (10--1000\,keV) and \textit{Insight}-HXMT (120--600\,keV), on public EP events up to December 2024. Their study reports flux upper limits for 12 EP events that were not visible by \textit{Fermi}/GBM. These upper limits, however, must be interpreted with caution: they are measured over a maximum timescale of 10\,s following the publicly available $T_0$, which can differ by up to tens of seconds from the final $T_{0,\,\rm X\text{-}ray}$ used in this work, and they do not cover the full duration of the burst as our GBM upper limits do (see Sec.\ref{sec:methodsGBM}). Consequently, they cannot be regarded as stringent constraints on the prompt $\gamma$-ray emission. Nevertheless, for the subset of EP events lacking any \textit{Fermi} constraint, these measurements currently represent the only available $\gamma$-ray upper limits; moreover, five of these 12 events have spectroscopic redshift measurements, making them particularly valuable for energetics and population studies. We therefore include all 12 upper 
limits in the final list of EP events without a confirmed $\gamma$-ray counterpart, converting them to the 10--1000\,keV energy range when necessary, in an effort to maximise the number of EP events with prompt emission constraints. 

Finally, for the 8 EP events still lacking any $\gamma$-ray constraint after the \textit{Fermi}-based analysis and the additional information described above, dedicated searches around the WXT trigger time and position were performed using SVOM/GRM and \textit{Swift}/BAT data. 
The \textit{Swift}/BAT targeted search was done using time-tagged event data obtained via GUANO \citep{2020ApJ...900...35T}, and the flux upper limits were computed using NITRATES \citep{2022ApJ...941..169D,2026arXiv260715130R}, while the search in \textit{SVOM}/GRM has been performed following the method described in \citealt{Cai2021}. This search yielded new constraints for two events, EP250125a (visible to both BAT-GUANO and \textit{SVOM}/GRM) and EP241111a (visible to \textit{SVOM}/GRM only), which had no associated signal. In the tables and figures presented in this work, we report the values derived from \textit{SVOM}/GRM, which yielded better constraints and whose characteristics are more similar to those of \textit{Fermi}/GBM. As a result, only 6 out of the 72 EP events in our sample remain without any $\gamma$-ray coverage.

\subsection{Search for counterparts in GBM}\label{sec:methodsGBM}

To identify potential \textit{Fermi}/GBM counterparts to EP/WXT transients, we first assessed the GBM visibility of each source during the corresponding EP/WXT duration. Despite GBM's 12 sodium iodide (NaI) and 2 bismuth germanate (BGO) detectors providing an instantaneous field of view covering the majority of the unocculted sky, visibility gaps arise from two reasons: Earth occultation or passages of the spacecraft through the South Atlantic Anomaly (SAA). Most events not visible to Fermi were affected by one of these two conditions for their entire duration. A few events were only partially visible during their $T_{\rm 100,X\text{-}ray}$, hence we evaluated visibility based on two criteria: an event was classified as not visible if (i) more than 50\% of its $T_{\rm 100,X\text{-}ray}$ fell within a visibility gap, and (ii) its peak time, $T_{\rm peak}$, also occurred during the same gap. For such events, no upper limit was derived. This occurred for 6 events; in total, 29 of the 49 events without a detected $\gamma$-ray counterpart and 15 of the 23 events with a detected $\gamma$-ray counterpart were visible to Fermi.

In parallel with the visibility assessment, we cross-matched the event list with the General Coordinates Network (GCN) Circulars to identify any reported $\gamma$-ray detections. For events not visible to \textit{Fermi}/GBM but associated with a GRB detected by another instrument (6 cases; see Sect.\,\ref{sec:GBM_results}), we report their $\gamma$-ray properties as derived from the corresponding instrument team's GCN Circular. When necessary, fluences and fluxes were converted to the 10--1000\,keV reference energy band by applying the spectral model reported in the respective GCN. For EP241113b, detected by the SVOM mission, the corresponding $\gamma$-ray flux values were obtained from private communication (Adrien et al., in prep.).

For events visible to \textit{Fermi}/GBM but lacking a detection in $\gamma$-ray emission, we run the Gamma-ray Targeted Search pipeline\footnote{\url{https://github.com/USRA-STI/gamma-ray-targeted-search}} \citep{Goldstein2019} to derive $3\sigma$ $\gamma$-ray flux upper limits. The Targeted Search was originally developed for follow-up searches of gravitational-wave (GW) signals \citep{Blackburn2015}. 
Here we employ a modified version adapted specifically for EP transients, incorporating three key modifications with respect to the standard configuration: (i) the search window is extended to $[-50.0,\,+250.0]$\,s relative to $T_0$ (compared to $[-30.0,\,+30.0]$\,s in the default GW configuration), which is based on the first associated events; (ii) the maximum duration timescale is increased to 32.768\,s (from 8.192\,s); and (iii) the spatial prior is set to a single point source position defined by the EP/WXT right ascension and declination (instead of the GW sky-map).
In cases where the event duration exceeded a single targeted search window, the search was repeated over consecutive intervals to ensure full temporal coverage of the emission episode. Three spectral templates were adopted for the upper limit computation: a \textit{soft} spectrum (Band function with $E_{\rm peak} = 70$\,keV, $\alpha = -1.9$, $\beta = -3.7$), a \textit{normal} spectrum (Band function with $E_{\rm peak} = 230$\,keV, $\alpha = -1.0$, $\beta = -2.3$), and a \textit{hard} spectrum (Comptonized function with $E_{\rm peak} = 1500$\,keV, $\alpha = -0.5$). For each EP event without a $\gamma$-ray counterpart, the search was performed twice: once over the full event duration $T_{\rm 100,X\text{-}ray}$, and once restricted to the peak time interval $T_{\rm peak,X\text{-}ray}$ (typically on a 1 s timescale; see Sec.~\ref{sec:GBM_results} for further discussion), to probe both the time-integrated and the peak $\gamma$-ray emission.

Crucially, the GBM $3\sigma$ flux upper limits were derived over the exact same time interval as the measured X-ray durations of each transient ($T_{\rm 100,X\text{-}ray}$ and $T_{\rm peak,X\text{-}ray}$), ensuring that the $\gamma$-ray and X-ray constraints are simultaneous. This approach is especially important for constraining the spectral properties of the transients, and for the time-averaged fluxes it also yields systematically deeper upper limits than those obtained over the generic search time window, due to the long duration of the X-ray emission. 
For comparison, in the analysis reported in \citet{Zhang2025_upperlimits} for GECAM and Insight-HXMT and \citet{Yadav2025} for GBM, the $\gamma$-ray upper limits are not constrained to be simultaneous with the X-ray emission. Furthermore, whereas \citet{Yadav2025} compared the GBM upper limits to the X-ray fluxes and durations reported in preliminary GCN Circulars, our analysis is based on the final, uniformly reduced EP/WXT measurements from the first EP extragalactic transient catalogue (Q. Wu et al., in prep), which provide refined trigger times $T_{0,\rm X-ray}$, fluxes and complete burst duration. This leads to non-negligible differences in both the $\gamma$- and the X-ray properties used for comparison.

\subsection{Search for counterparts in LAT}\label{sec:LATselection}

To identify potential GeV counterparts to EP/WXT transients, we first assessed whether \textit{Fermi}/LAT was observing each source during the time of the EP/WXT detection. Two geometric conditions during the period have been checked: the LAT off-axis angle ($\theta$: angle between the source and the LAT boresight) and the source zenith angle 
($Zn$: angle between the source and the local zenith of the LAT; \citealt{Atwood2009_LAT}).

The sensitivity of LAT drops sharply for off-axis angles larger than \(70^\circ\), due to the steep decrease in effective area\footnote{\url{https://www.to.infn.it/~maldera/Fermi-LAT/test_LATperf/lat_Performance.htm}}. 
Additionally, zenith angles greater than \(100^\circ\) can introduce contamination from $\gamma$-rays produced in the Earth’s limb. 
Therefore, we define the time intervals during which both conditions are satisfied ($\theta$\(< 70^\circ\) and $Zn$\(< 100^\circ\)) as the periods when LAT was observing the source.
For the transients continuously visible by LAT throughout the EP/WXT duration, we fully analyzed LAT data.
When the LAT visibility interval does not fully overlap with the EP/WXT duration window, we select sources for dedicated LAT analysis using two criteria. We retain cases where the LAT coverage exceeds 50\% of the EP/WXT duration, and, for lower-overlap cases, we retain only those in which the EP/WXT light-curve peak time is covered. This allows us to include sources for which LAT samples either a substantial fraction of the transient or its peak emission. Two analyzed transients have LAT coverage exceeding 50\% of the EP/WXT duration even though the EP/WXT peak itself is not covered. The LAT--WXT coverage information for all sources analyzed with LAT is summarized in Table~\ref{Table_LAT}.

Applying these selection criteria, among the initial 72 EP/WXT transients, only 12 were found to satisfy the LAT visibility criteria and were subsequently analysed for the GeV counterparts. 
For each of these 12 EP events visible by LAT, we performed an unbinned likelihood analysis of \textit{Fermi}/LAT data within the energy range of 0.1–1 GeV, using the \texttt{GTBURST} software \footnote{\url{https://fermi.gsfc.nasa.gov/ssc/data/analysis/scitools/gtburst.html}} from the Fermi Science Tool. A region of interest of $12^{\circ}$ around the source location as determined by the EP/WXT trigger location has been chosen. We used the \texttt{P8R3$\_$TRANSIENT020} event class that is appropriate for analyzing transient sources with short exposures, along with the relevant instrument response functions. The analysis included isotropic particle background (\texttt{`isotr template'} in \texttt{GTBURST}), and galactic and extragalactic high-energy components from the Fermi Fourth Catalog (4FGL), all with fixed normalization (\texttt{`template (fixed norm)'}). A power-law spectral model (\texttt{`powerlaw2'} was fitted and for nondetection cases, 3$\sigma$ upper limits are derived assuming three different photon indices, $\Gamma_{\rm LAT}=-1.5$, $-2.0$, and $-2.5$, corresponding to hard, intermediate, and soft spectra, respectively. 

\section{Results}\label{sec:results}
\subsection{GBM}\label{sec:GBM_results}

A summary of the search is reported in Table \ref{tab:EP_stats}. Out of the 72 events, 44 (61\%) were visible by Fermi/GBM. We make the results of our \textit{Fermi} Targeted Search analysis publicly available at the \href{https://fermi.gsfc.nasa.gov/ssc/data/access/gbm/targeted_search/}{ Fermi GBM Online Targeted Search webpage}. In the following, we describe the results of our search, according to the presence or absence of a $\gamma$-ray detection. 

\begin{table}[htbp]
\small
\centering
\caption{Summary of the \textit{Einstein Probe} (EP) unknown transient sources detected between February 2024 and February 2025, classified according to the presence of a $\gamma$-ray counterpart, their visibility by $\gamma$-ray satellites and redshift measurements.}
\begin{tabular}{lc}
\toprule
Category & Number of sources \\
\midrule
\textbf{GRB counterpart} & \textbf{23 (32\%)} \\
\hspace{1em} visible by \textit{Fermi}/GBM & \hspace{1em} 15 \\[3pt]
\hspace{1em} redshift known & \hspace{1em} 9 \\[3pt]
\textbf{No $\gamma$-ray counterpart} & \textbf{49 (68\%)} 
\\[3pt]
\hspace{1em}  visible by $\gamma$-ray satellites yet undetected & \hspace{1em} 43 \\[3pt]
\hspace{4em}  of which visible by \textit{Fermi}/GBM & \hspace{2em} 29 \\[3pt] 
\hspace{1em} redshift known & \hspace{1em} 14 \\[3pt] 
\midrule
\textbf{Total} &  \textbf{72} \\ 
\bottomrule
\end{tabular}
\label{tab:EP_stats}
\end{table}

\subsubsection{Events with a $\gamma$-ray counterpart}
Twenty-three sources (32\% of the sample) have a $\gamma$-ray counterpart, of which 5 have been identified offline by the Fermi Targeted Search pipeline used in this work (i.e. did not trigger Fermi/GBM on board). 
Among these 23 GRBs, 8 occurred outside the visibility of Fermi/GBM (either Earth-occulted or during SAA), but a $\gamma$-ray counterpart was independently detected by other missions (see Table \ref{tab:Gamma_counterparts}). As described in Sec.\ref{sec:sampleselection}, two events, EP240202a and EP240703b, have been identified by a targeted search in Konus-Wind data (Wu et al. 2026, in preparation). However, the association for these two events remains tentative, and the corresponding $\gamma$-ray properties are not yet available; we therefore exclude them from the comparison analysis presented in this work.
Nine of the 23 GRB-associated events (39\% of the sample) have a spectroscopic redshift measurement. In these cases, we additionally compute the isotropic-equivalent luminosity $L_{\rm iso}$ and energy $E_{\rm iso}$. For all these GRB-associated events, we report their $\gamma$-ray properties (time-averaged $\gamma$-ray flux, peak flux and $T_{90}$) and the redshift, when available, in Table \ref{tab:Gamma_counterparts}. 

\subsubsection{Events without a $\gamma$-ray counterpart}
The remaining 49 sources (68\% of the sample) have no reported $\gamma$-ray counterpart; among them, 29 were visible to GBM, for which we derived $3\sigma$ upper limits on the 10-1000 keV flux using the three spectral templates described in Sect.~\ref{sec:methodsGBM} (soft, normal, and hard). 
Upper limits obtained with the normal (hard) template are on average a factor of $\sim$1.5 ($\sim$3) higher than those derived with the soft template.
Since the majority of the detected counterparts are best fitted by the Targeted Search with the \textit{normal} template, and given that this template yields intermediate upper limits values among the three spectral assumptions, we adopt the \textit{normal} template as the default one throughout the paper. The full list of upper limits derived for all three spectral templates is reported in Table \ref{tab:NoGamma_counterparts}.

The remaining 20 sources were not visible to GBM. As described in Sec.\ref{sec:sampleselection}, we additionally checked their visibility from \textit{Swift}/BAT, finding that 7 of them were within the BAT field of view during the event, yet no associated upper limit or detection was reported. However, 12 of these \textit{Fermi}-unconstrained events were visible by GECAM and \textit{Insight}-HXMT, hence we supplement our analysis with the upper limits reported by \citet{Zhang2025_upperlimits}, converted to the 10--1000\,keV energy range when necessary. These are included in Table~\ref{tab:NoGamma_counterparts}.  
After accounting for them, only eight events remain with no coverage, namely that were not visible to either \textit{Fermi}/GBM or \textit{Swift}/BAT, and for which, crucially, no $\gamma$-ray detection has been reported by any other high-energy mission either. We therefore treat these events as having no identified $\gamma$-ray counterpart throughout this work.

Redshift information is available for a total of 14 (29\%) of these 49 non-GRB events. We note that, in relative terms, redshifts are more frequently measured for events with a confirmed $\gamma$-ray counterpart. 
However, the larger absolute number of redshift measurements among non-GRB events likely reflects the accurate and rapid X-ray localisation provided by EP/WXT, which enables fast optical follow-up independently of a $\gamma$-ray trigger. This is despite the fact that, for some events, the delay in alerting the community and obtaining the X-ray afterglow position with EP/FXT has been quite long.

\subsubsection{X-ray to $\gamma$-ray flux comparison}
For each event, it is interesting to compare the fluxes measured in the X-ray and $\gamma$-ray energy bands. 
One of the key results of the \textit{Einstein Probe} mission is that the total burst duration observed in X-rays is substantially longer than that observed in $\gamma$-rays, consistent with earlier indications from BeppoSAX (\citealt{Frontera2000,Frontera2004,Piro2005} and more recently in \citealt{BeppoSAX2026}), although we note the rather different energy bands of WXT (0.5-4 keV) and BeppoSAX Wide Field Camera (WFC; 2-30 keV). For the events analysed in this work, the X-ray duration $T_{\rm 100,X\text{-}ray}$ is a factor of $\gtrsim 5$ longer than the $\gamma$-ray duration $T_{90}$, with a median ratio of $\sim$12. Although part of the difference may reflect the use of different duration definitions ($T_{\rm 100,X\text{-}ray}$ versus $T_{90}$), the much longer X-ray durations indicate that this effect alone cannot account for the observed discrepancy, implying that the X-ray emission generally persists significantly longer than the detectable prompt $\gamma$-ray emission. As a consequence, a direct comparison of the X-ray and $\gamma$-ray fluxes measured over their respective durations would not be simultaneous. This motivated us, for those EP sources with a $\gamma$-ray counterpart, to compute the X-ray flux restricted to the shorter $\gamma$-ray emission time interval $T_{90}$, allowing for a simultaneous comparison. This was possible for 18 out of 23 GRB-associated events.  For the remaining 5, no simultaneous WXT flux could be derived: for 3 of them, listed in Table \ref{tab:Gamma_counterparts_butnoEPdetected}, EP/WXT observations started $\sim$minutes after the $\gamma$-ray trigger, precluding simultaneous coverage and suggesting that WXT may have triggered on the afterglow rather than on the prompt emission; the other 2 are the tentative Konus-Wind associations discussed above (EP240202a and EP240703b), for which the corresponding $\gamma$-ray properties are not yet available.
\\

\begin{figure}
   \centering
   \includegraphics[width=\hsize]{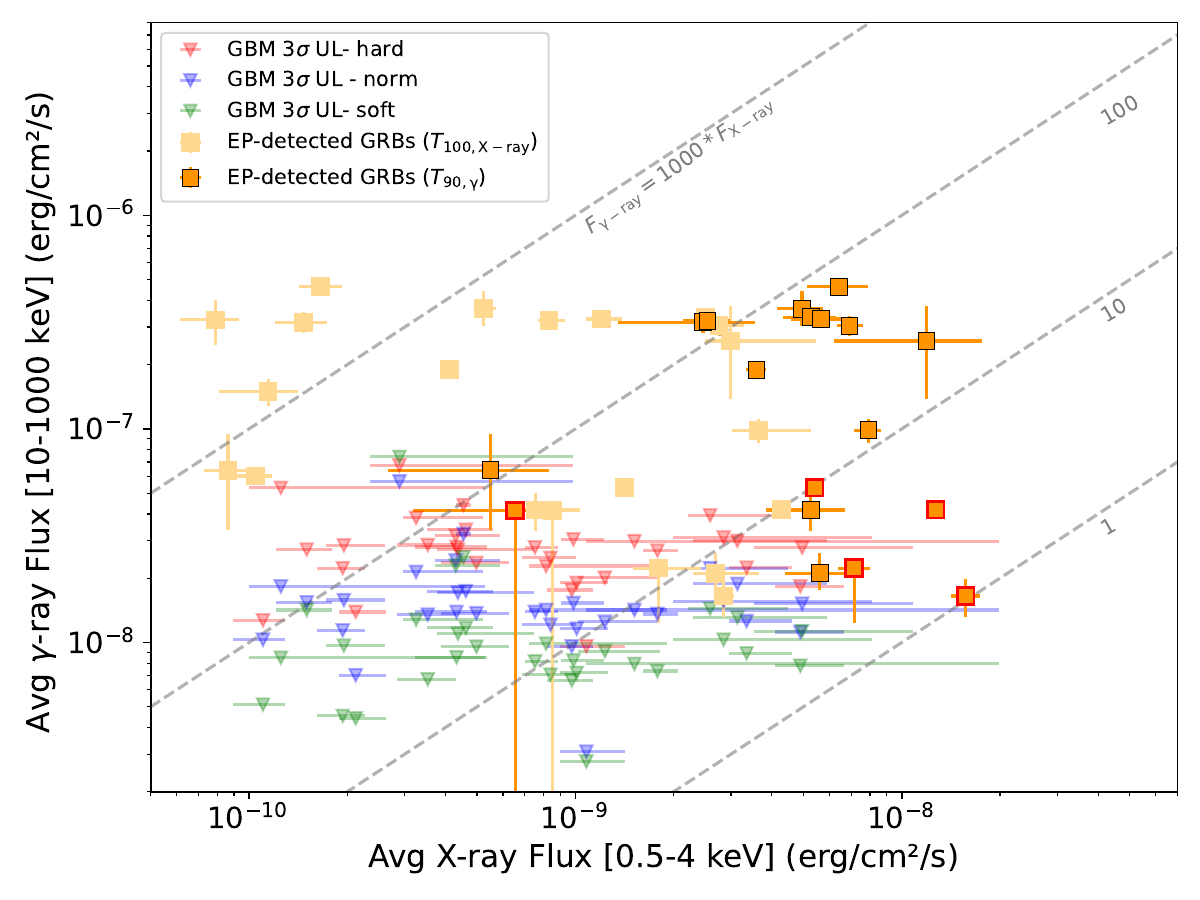}
      \caption{Comparison between average fluxes in $\gamma$-ray [10-1000 keV] and in X-ray energy bands [0.5-4 keV] for the sample of EP events analysed in this work. EP sources with a $\gamma$-ray counterparts are represented by squares: light beige when the X-ray flux is measured over the total X-ray duration ($T_{100, \rm X-ray}$), and orange when it is measured simultaneously to the $\gamma$-ray emission (i.e. during $T_{90, \rm \gamma}$). Events that did not trigger \textit{Fermi}/GBM but were recovered via the Targeted Search are marked with a red contour. In case of non-detections, the GBM 3-$\sigma$ upper limits simultaneous to $T_{100, \rm X-ray}$ are shown as downward arrows, colour-coded according to the three spectral shapes assumed (green: soft; blue: normal; red: hard; see text for details). Purple downward arrows represent the 10s timescale upper limit derived in \citealt{Zhang2025_upperlimits} using either GECAM or Insight-HXMT (not simultaneous to the X-ray flux). Grey dashed lines indicate constant ratios between the $\gamma$-ray and the X-ray fluxes.}
         \label{fig:cfr_GBMavgflux}
\end{figure}

The distribution of X-ray and $\gamma$-ray fluxes is shown in Figure \ref{fig:cfr_GBMavgflux}. In case of non-detections, the corresponding GBM 3$\sigma$ upper limits are shown as downward triangles (green, blue and red) corresponding to the three spectral shapes assumed (soft, normal and hard). They have been computed over the total burst duration in EP/WXT ($T_{100, \rm X-ray}$), hence they are simultaneous to the X-ray emission. The upper limits derived in \citealt{Zhang2025_upperlimits} using either GECAM or Insight-HXMT are represented by purple downward arrows (on a 10s timescale, not simultaneous with the X-ray flux). 
In case of detections, events with a GRB association are represented by squares: light beige squares indicate the X-ray flux computed over $T_{\rm 100,\,X-ray}$, while orange squares indicate the X-ray flux computed over $T_{\rm 90,\,\gamma}$, i.e.\ simultaneous with the $\gamma$-ray emission. Those whose $\gamma$-ray counterparts have been found via the Targeted Search are represented by a red contour.

When considering the X-ray flux integrated over the full duration of the event ($T_{100,\mathrm{X-ray}}$), no clear separation is observed between EP sources with a detected $\gamma$-ray counterpart (light beige squares) and those without it (downward arrows). Although the two populations clearly differ in their $\gamma$-ray fluxes, they span a remarkable similar range of X-ray fluxes. A Kolmogorov-Smirnov (KS) test does not reject the null hypothesis that EP sources with and without detected $\gamma$-ray counterparts are drawn from the same parent population ($p_{\rm KS}=0.60$). If two EP events have similar X-ray fluxes, one could be detected in $\gamma$-ray while the other one not. 
This indicates that the average X-ray brightness alone is not predictive of the presence of detectable $\gamma$-ray emission. This is an important finding offered by the first year EP sample and it suggests that additional factors, such as spectral properties and lightcurve evolution may determine whether the high-energy counterpart is detectable or not.  

Perhaps not surprisingly, when the X-ray flux is computed over the time interval simultaneous with the $\gamma$-ray emission ($T_{90,\gamma}$; orange squares), the X-ray fluxes shift systematically toward higher values. This suggests that $\gamma$-ray emission is preferentially detected during episodes of enhanced X-ray activity, pointing towards a common triggering condition. 
We find that the ratio between the simultaneous $\gamma$-ray and X-ray fluxes typically lies in the range $1 \lesssim F^{\mathrm{avg}}_{\gamma}/F^{\mathrm{avg}}_{\mathrm{X-ray}} \lesssim 100$, indicated by the grey dashed lines in Figure~\ref{fig:cfr_GBMavgflux}. A KS test between the distributions of simultaneous X-ray fluxes for GRB-associated events and the X-ray fluxes of non-GRB events yields $p_{\rm KS}<10^{-4}$, supporting the interpretation that $\gamma$-ray emission is linked to the brightest phases of the X-ray activity.
While the $\gamma$-ray emission selects the phases with higher X-ray fluxes, this does not exclude the possibility that it might be present during fainter phases of the X-ray emission, and it remains below the sensitivity threshold. Indeed, a simple comparison of typical sensitivities between WXT ($\sim 10^{-11} \rm erg/cm^2/s$, for a 1 ks exposure time) and GBM ($\sim 10^{-8} \rm erg/cm^2/s$) shows that WXT can probe much fainter emission episodes in the 0.5-4 keV band than those accessible to GBM in 10-1000 keV.
In this context, the flux ratio becomes particularly informative, as the simultaneity of the emission allows the use of X-ray and $\gamma$-ray measurements for building a spectral energy distribution, providing constraints on the spectral shape.
This will be further discussed in Sec.\ref{sec:discussion_SED}.\\

\begin{figure}
   \centering
   \includegraphics[width=\hsize]{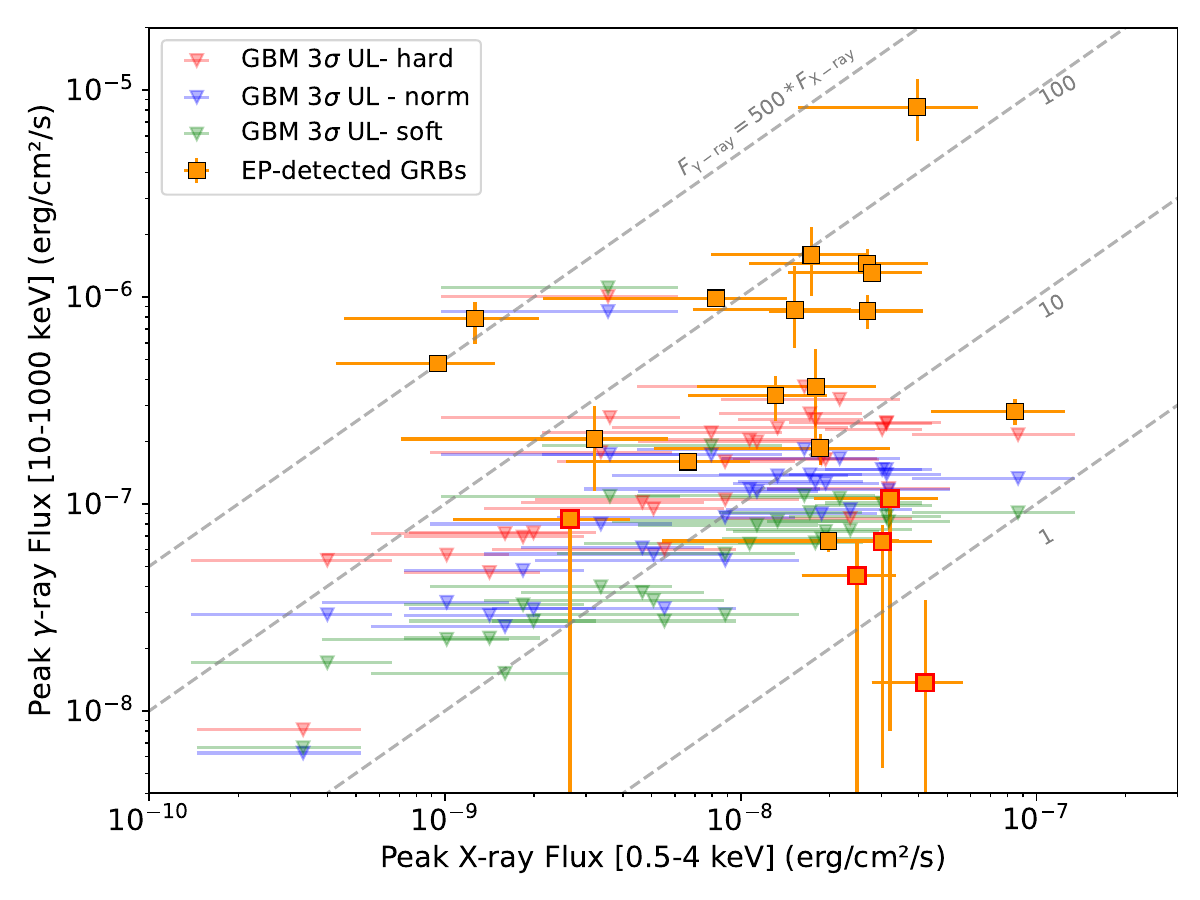}
      \caption{Same as Figure~\ref{fig:cfr_GBMavgflux}, but with peak fluxes computed at the respective peak times of the X-ray and $\gamma$-ray lightcurve. }
         \label{fig:cfr_GBMpeakflux}
\end{figure}

Similarly to Figure \ref{fig:cfr_GBMavgflux}, Figure \ref{fig:cfr_GBMpeakflux} shows the comparison between the derived peak X-ray flux and the peak $\gamma$-ray flux for each event, with the same color-code.
For GRB-associated events that triggered \textit{Fermi}/GBM onboard, we used the peak flux values reported in the official GBM catalogue. For sub-threshold events identified through the Targeted Search, we performed an independent spectral analysis to derive the peak $\gamma$-ray flux, modeling the spectra with a cutoff power-law (CPL) function, which was found to provide an adequate description for the majority of the faint GBM counterparts detected in our sample. For the GRB-associated events that were not visible by Fermi, we adopted the peak flux reported via GCN by other high-energy missions (mainly Konus-Wind), which we converted to the 10-1000 keV energy band assuming the reported best-fit model (mainly CPL). Conversely to the time-averaged case, no information at peak times are reported in  \citet{Zhang2025_upperlimits}, hence no additional peak fluxes upper limits are available to consider. 
We note that, in most cases, the peak fluxes were computed over a 1\,s timescale; however, for a few faint events, we measured the flux over the shortest time interval allowed by the available photon statistics in the EP/WXT data, as well as by the temporal resolution of the non-trigger continuous data from \textit{Fermi}/GBM (typically 4\,s).

Events with a $\gamma$-ray counterpart tend to exhibit higher X-ray peak fluxes than those without $\gamma$-ray detections, with a median WXT peak flux about twice as high, although the difference is not statistically significant in our sample (KS test yields $p=0.13$). This hints at a preferential association between $\gamma$-ray detectability and brighter X-ray episodes, supporting the finding described in the previous section. The ratio between $\gamma$-ray and X-ray peak fluxes is found to be in the range $1 \lesssim F^{\mathrm{peak}}_{\gamma}/F^{\mathrm{peak}}_{\mathrm{X-ray}} \lesssim 500$.

A caveat to this comparison is that, while the upper limits on the peak flux are simultaneous by construction (as in Fig.~\ref{fig:cfr_GBMavgflux}), the situation for detected events is more complex, as the exact 1s peak intervals in the X-ray and $\gamma$-ray bands do not always coincide for individual events. This is likely driven by intrinsic evolution of the light curve rather than by the limitations in photon statistics or temporal resolution mentioned above. As a consequence, unlike the average-flux comparison, the X-ray peak fluxes reported in Fig.~\ref{fig:cfr_GBMpeakflux} are not strictly simultaneous with the $\gamma$-ray peak fluxes. For events where the X-ray and $\gamma$-ray peaks are offset, we checked the signal-to-noise ratio in the GBM data at the time of the X-ray peak and, unfortunately, found it insufficient to derive a reliable $\gamma$-ray flux measurement, preventing a simultaneous constraint.

\subsection{LAT}
\begin{figure}
   \centering
   \includegraphics[width=\hsize]{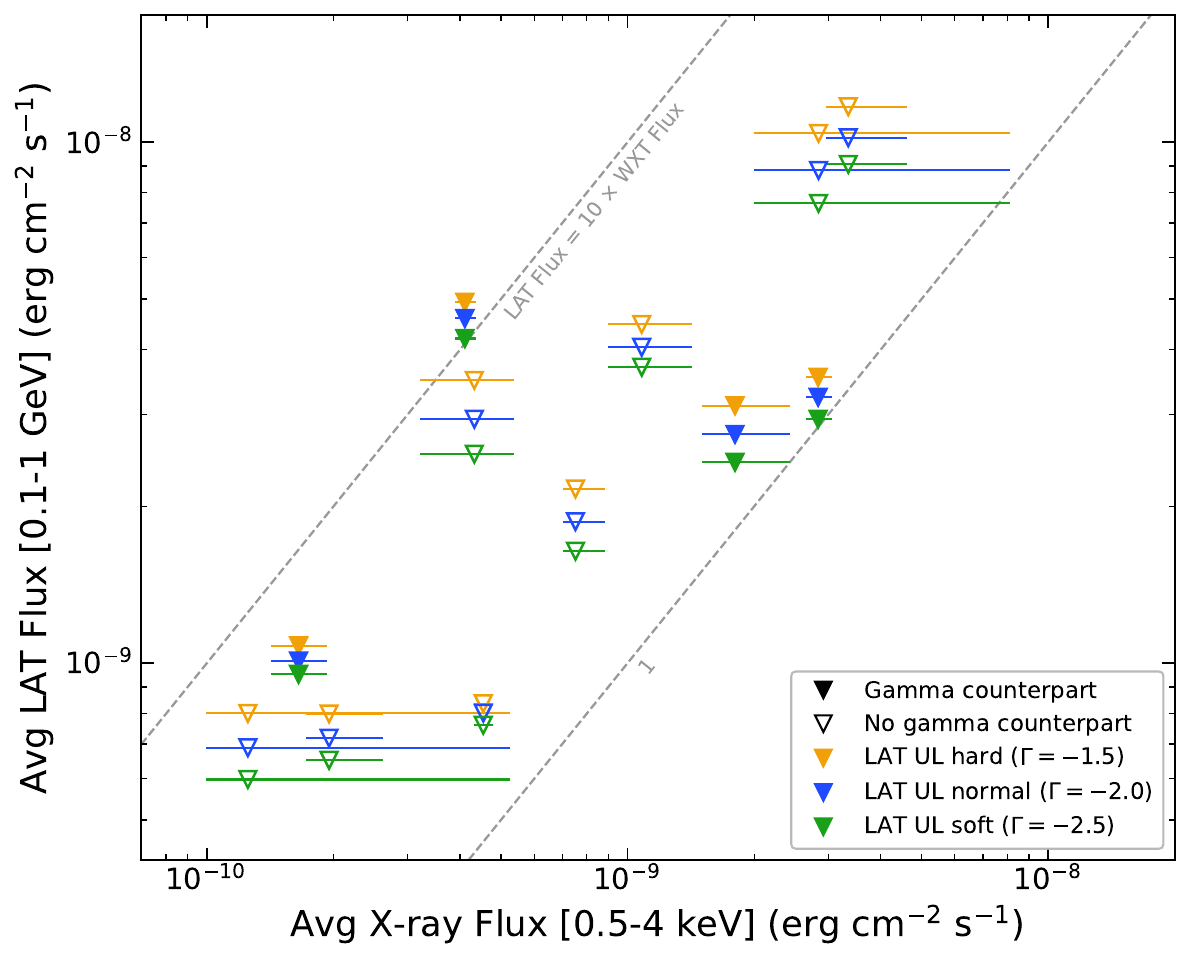}
      \caption{Time-averaged flux in the GeV band compared to the time-averaged flux in the EP/WXT band for the total duration (T$_{100, \rm X-ray}$) of the transients selected in this work (see Sect.\ref{sec:LATselection}). 
      The GeV upper-limits are estimated using three spectral indices, hard: $-1.5$ (orange triangle), normal: $-2.0$ (blue triangle), and soft: $-2.5$ (green triangle). The two dashed lines indicate the equality line between the fluxes and 10 times the flux of EP/WXT. In addition, the empty triangles indicate the transients without GRB counterparts.}
         \label{fig:cfr_LATavgflux}
\end{figure}

Out of the 72 EP transients considered in this work, 12 satisfied the LAT selection criteria described in Section~\ref{sec:LATselection}. The upper limits of the resulting GeV flux for these 12 sources are presented in Table~\ref{Table_LAT}, and their time-averaged X-ray and LAT fluxes are plotted in Figure~\ref{fig:cfr_LATavgflux}. Four of the 12 EP transients are associated with a $\gamma$-ray counterpart in the 10-1000 keV energy range. 

For these 12 events, we initially searched for high-energy emission in the 0.1--100 GeV LAT energy range during the fraction of the T$_{100, X-ray}$ interval covered by LAT. Among the selected sources, only EP241025a showed evidence of GeV emission, with the detection of a $\sim$3 GeV photon with an association probability of 99.94\% within the search time interval. This is a confirmed GRB in Fermi/GBM. The corresponding likelihood analysis\footnote{The test statistic is given by ${\rm TS} = 2\left(\log \mathcal{L}_1 - \log \mathcal{L}_0\right)$, where $\mathcal{L}_1$ and $\mathcal{L}_0$ are the likelihoods of the data with and without the source, respectively. Larger values of TS indicate stronger evidence for the presence of a source. $\sqrt{\rm TS}$ approximately corresponds to the detection significance in Gaussian sigma.} gives ${\rm TS}=21$, corresponding to an approximate pre-trial significance of $\sim 4.6\sigma$ ($>4\sigma$ post-trial\footnote{Since the spectral index and normalization are treated as free parameters, we estimate the detection significance using Wilks' theorem \citep{d543aecb-cd73-36d5-9101-f08a74f8e8c6} with a $\chi^2$ distribution with two degrees of freedom. This gives a tail probability
$p = P\left(\chi^2_{2} \geq 21\right) \simeq 2.75\times 10^{-5}$,
corresponding to an equivalent one-sided Gaussian significance of $\simeq 4.03\sigma$.}) in the 0.1--100 GeV energy range.

Although this excess is suggestive, it remains below the standard LAT detection threshold of TS = 25, which is generally adopted for a detection. Given the absence of significant detections, we restricted our analysis to the 0.1--1 GeV energy range and derived 3$\sigma$ flux upper limits over the fraction of the T$_{100, \rm X-ray}$ for which the off-axis angle $\theta$\(< 70^\circ\), and the zenith angle $Zn$\(< 100^\circ\). The resulting average LAT upper limits are compared with the average WXT fluxes in Fig.~\ref{fig:cfr_LATavgflux}. For the analyzed sample, the ratio between the GeV and X-ray fluxes is $F^{\rm avg}_{0.1\text{-}1\,{\rm GeV}}/F^{\rm avg}_{0.5\text{-}4\,{\rm keV}} \lesssim 10$. 
For comparison, only $\sim$8\% of \textit{Swift}/BAT GRBs were detected by the \textit{Fermi}/LAT during the period covered by the second LAT GRB catalogue \citep{Tiwari2025}. Given this low detection fraction, the absence of LAT counterparts to EP transients is not statistically unexpected and is consistent with these events being intrinsically less energetic than classical GRBs.

\begin{figure}
   \centering
   \includegraphics[width=\hsize]{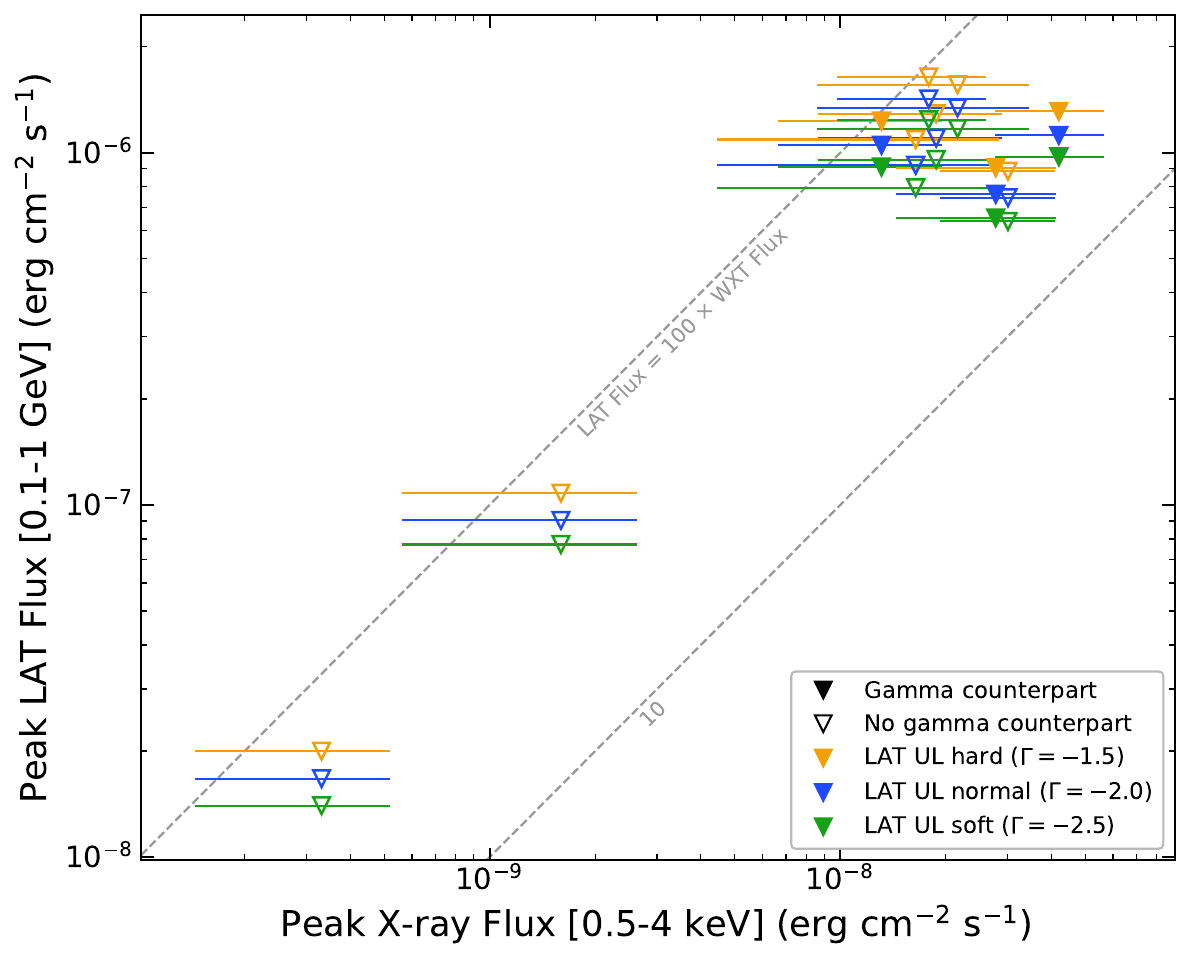}
      \caption{Same as Fig. \ref{fig:cfr_LATavgflux}, but with the EP/WXT and Fermi/LAT fluxes derived at peak times. The two dashed lines represent the 10 and 100 times flux ratio between LAT and WXT energy bands.}
         \label{fig:cfr_LATpeakflux}
\end{figure}

We also derived LAT upper limits over the time interval corresponding to the peak of the X-ray emission. This analysis could be carried out for 10 of the 12 sources, while the remaining two were outside the LAT field of view during the X-ray peak interval. Figure~\ref{fig:cfr_LATpeakflux} compares the LAT upper limits obtained during the X-ray peak time with the corresponding peak X-ray fluxes. As expected, the upper limits derived over the shorter peak intervals are less constraining than those obtained over the full X-ray duration, owing to the reduced LAT sensitivity for shorter integration times. The obtained limits indicate that any GeV emission associated with the X-ray peak is constrained to remain below approximately two orders of magnitude of the corresponding X-ray flux. We report in Table~\ref{Table_LAT} all information about the 12 EP transients analysed with LAT.

\subsection{X-ray to GeV spectral energy distribution}\label{sec:SED_results}

\begin{figure*}
    \centering
    \begin{subfigure}{0.497\textwidth}
        \includegraphics[width=\linewidth]{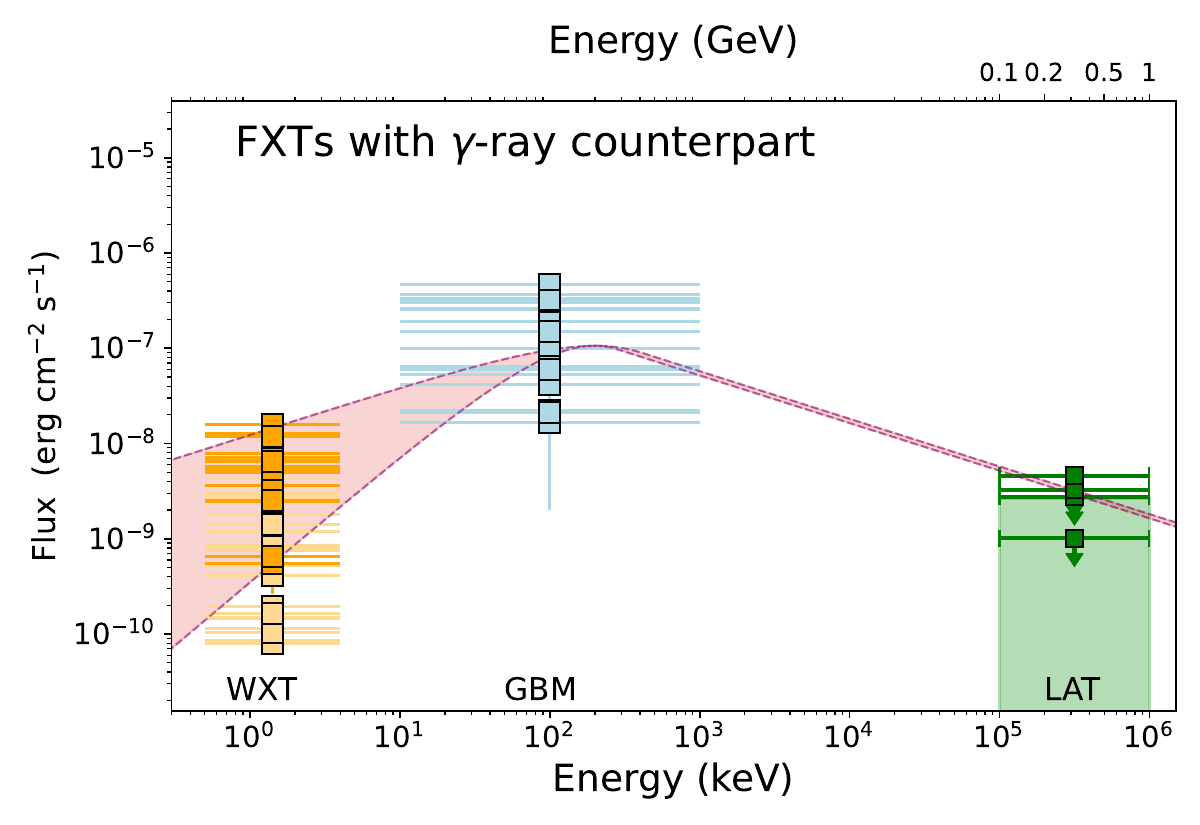}
    \end{subfigure}
    \hfill
    \begin{subfigure}{0.497\textwidth}
        \includegraphics[width=\linewidth]{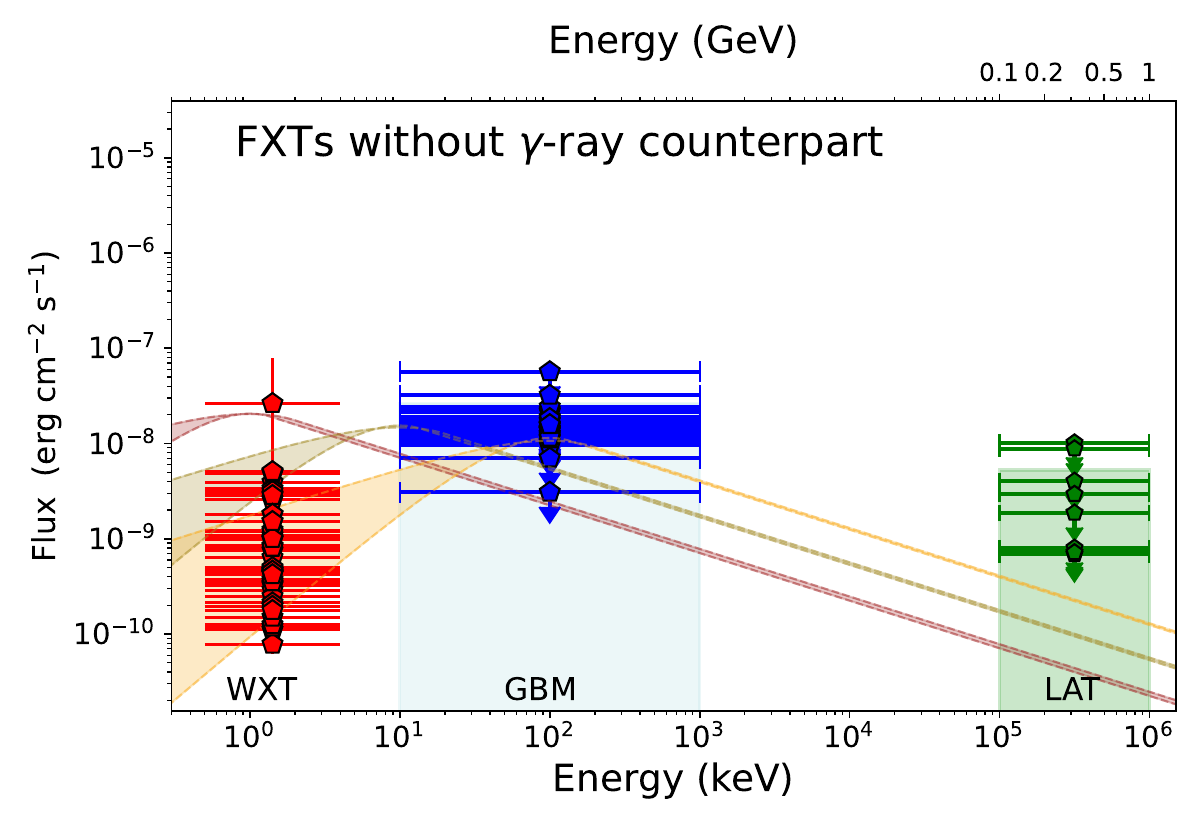}
    \end{subfigure}
    \caption{Broadband spectral energy distribution from X-rays to GeV energies for transients detected during the first year of \textit{Einstein Probe}. \textit{Left panel:} SED for the EP transients with a $\gamma$-ray counterpart. The time-averaged (simultaneous) EP/WXT fluxes in the 0.5--4 keV band are shown as beige (orange) squares (see Sec.\ref{sec:discussion_comparisonFermiGRB} for the details), together with the corresponding average flux measurements in the 10--1000 keV band (light blue squares). For events within the \textit{Fermi}/LAT field of view, the corresponding 0.1--1 GeV upper limits are also reported (green downward arrows). A representative Band function with $E_{\rm peak}=200$ keV reproducing the average observed integrated fluxes is shown with a dashed purple line, with shaded regions indicating $-1.5 \leq \alpha \leq -0.67$. For higher $E_{\rm peak}$ and soft low-energy slopes, the extrapolated spectra are likely to exceed the observed WXT fluxes, suggesting that a low-energy spectral break may be required in some events.
    \textit{Right panel:} EP transients without a detected $\gamma$-ray counterpart. Average EP/WXT fluxes in the 0.5--4 keV band are shown as red pentagons. For sources visible by $\gamma$-ray instruments, the derived upper limits in the 10--1000 keV band are reported as blue downward arrows, while for events visible by \textit{Fermi}/LAT, the corresponding 0.1--1 GeV upper limits are shown as green downward arrows. As in the left panel, dashed lines show representative Band functions reproducing the observed integrated fluxes, here with $E_{\rm peak}=1$, 10, and 100 keV (brown, ochre, and orange, respectively).}
    \label{fig:SED}
\end{figure*}

Using the $\gamma$-ray flux measurements and upper limits presented in the previous sections, we construct the broadband spectral energy distributions (SEDs) from X-rays to GeV energies for both samples of EP transients, those associated with a $\gamma$-ray counterpart and those without it. The resulting SEDs are shown in Figure~\ref{fig:SED}.

For EP transients associated with a $\gamma$-ray counterpart (left panel), we report both the total (light beige squares) and the simultaneous (orange squares) time-averaged X-ray flux measured by EP/WXT in the 0.5--4 keV energy range, together with the corresponding average $\gamma$-ray flux measured in the 10--1000 keV band (light blue squares). For events within the \textit{Fermi}/LAT field of view, the corresponding 0.1--1 GeV flux upper limits are shown as green downward arrows. These broadband SEDs are presented for comparative purposes only: no independent or joint spectral fitting across the X-ray and $\gamma$-ray datasets has been performed. Properly assessing the spectral connection between the X-ray and $\gamma$-ray emission and deriving a common photon index, key for characterising the prompt emission mechanism, would require a joint fit between the two datasets, which is beyond the scope of this work.

For EP transients without an identified $\gamma$-ray counterpart (right panel), we show the time-averaged EP/WXT X-ray fluxes in the 0.5--4 keV band (red pentagons). For events observable by GBM, we report the derived upper limits in the 10--1000 keV band, assuming the normal spectral template described in Section~\ref{sec:methodsGBM}, displayed as blue downward arrows. Analogously, sources visible by LAT are marked with green downward arrows corresponding to the derived 0.1--1 GeV upper limits.

The broadband representation in Figure~\ref{fig:SED} extends over approximately six orders of magnitude in energy and provides a complementary visualisation of the relationship between X-ray and $\gamma$-ray emission displayed in Figure \ref{fig:cfr_GBMavgflux} and Figure \ref{fig:cfr_LATavgflux}. 
For comparison, we overplotted representative Band functions whose integrated fluxes in WXT, GBM and LAT bands would be compatible with the derived flux measurements or upper limits. In particular, in the left panel, we considered a Band function with a peak energy $E_{\rm peak}$=  200 keV (dashed purple lines). The shaded regions indicate the range of low-energy photon indices $\alpha$ between -0.67 and -1.5. It is likely that, for higher $E_{\rm peak}$ and soft low-energy slopes, the extrapolated spectra can exceed the observed WXT fluxes, suggesting that a low-energy spectral break as those found in e.g. \citet{Oganesyan2018, Ravasio2019} may be required in some events. 
In the right panel, we considered a Band function with peak energies $E_{\rm peak}$= 1, 10, and 100 keV (shown as dashed orange, ochre, and brown lines, respectively), all of which remain consistent with our upper limits. We note that the 'normal' template ($E_{\rm peak}$= 200 keV) is only a representative spectral shape used to compute the flux upper limits; as illustrated in the right panel of the figure, softer spectral shapes remains entirely plausible for these events.
We will discuss the physical implications of these broadband limits in Section~\ref{sec:discussion_SED} and Section~\ref{sec:discussion_SSC}.

\section{Discussion}\label{sec:discussion}

\subsection{Constraints on the spectrum}\label{sec:discussion_SED}

\begin{figure}
   \centering
   \includegraphics[width=0.5\textwidth]{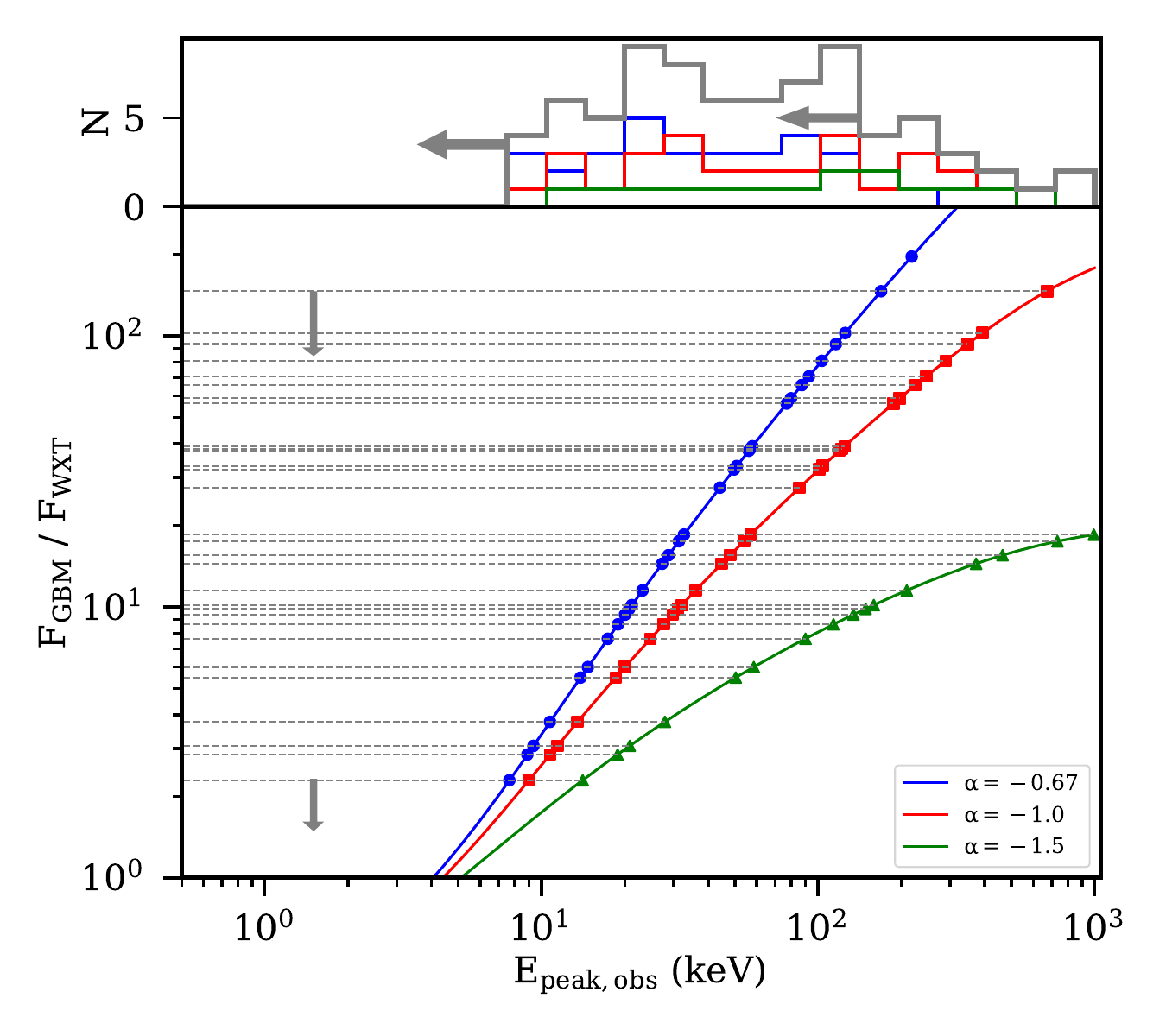}
   \caption{Flux ratio between the GBM [10-1000 keV] and WXT [0.5-4 keV] energy bands as a function of the peak energy, $E_{\rm peak}$, for a Band spectral model with fixed $\beta=-2.3$ and different low-energy photon indices: $\alpha=-0.67$ (blue curve), $-1$ (red curve), and $-1.5$ (green curve). 
   Horizontal grey dashed lines with downward arrows represent different ratios between the measured X-ray flux for EP events without a $\gamma$-ray counterpart and the corresponding \textit{Fermi}/GBM upper limits. }\label{fig:fluxratios}
\end{figure}

The different instrument sensitivities in different energy bands may prevent a proper assessment on the intrinsic lack of simultaneous emission in $\gamma$-rays, which might be simply too faint for current high-energy instruments. Indeed, some of the EP events without detected $\gamma$-ray counterparts may have a $\gamma$-ray emission, but its spectral shape, in terms of photon indices and peak energy, affects its detectability in $\gamma$-rays. For example, a Band spectrum with $E_{\rm peak} \sim$ 300 keV could have a comparable X-ray flux [0.5–4 keV] to that of a softer event with $E_{\rm peak} \sim$ 5 keV, while the flux in $\gamma$-ray will be $\sim$2 orders of magnitude fainter in the latter case and may remain below the sensitivity threshold of \textit{Fermi}/GBM\footnote{The exact ratio varies with the assumed low-energy photon index $\alpha$, but the order-of-magnitude contrast is preserved.}. The flux ratio  $F_{\gamma}/F_{\rm X}$ therefore carries some diagnostic value: while for some events the $\gamma$-ray upper limit is not sufficiently deep to constrain the high-energy emission, for others this ratio approaches unity, providing the most stringent constraints on the spectral properties of the source.
To investigate the maximum value of peak energy that is allowed by the current upper limits derived from Fermi/GBM, we simulated a Band function with a fixed $\beta$=-2.3 (the same used in the spectral template), and computed the flux ratio between 10-1000 keV and 0.5-4 keV energy bands as a function of the peak energy, and repeated this for three different low-energy photon indices $\alpha$ (-0.67, -1, -1.5). Since we are interested in the flux ratio, the normalization of the spectrum is irrelevant. The results are shown in Figure \ref{fig:fluxratios}. 
As expected, for high ratios ($\gtrsim$ 10), the upperlimits have less constraining power, inferring $E_{\rm peak} \lesssim 500$ keV. The stringest constraints on the position of the peak energy comes from those events with a $\gamma$-ray flux upper limit a factor $\lesssim$ 1-10 times the X-ray flux, where we can derive $E_{\rm peak} \lesssim 30$ keV for $\alpha$=-1 or -0.67 and $\lesssim 100$ keV for $\alpha$=-1.5. 

\begin{figure}
   \centering
\includegraphics[width=0.545\textwidth]{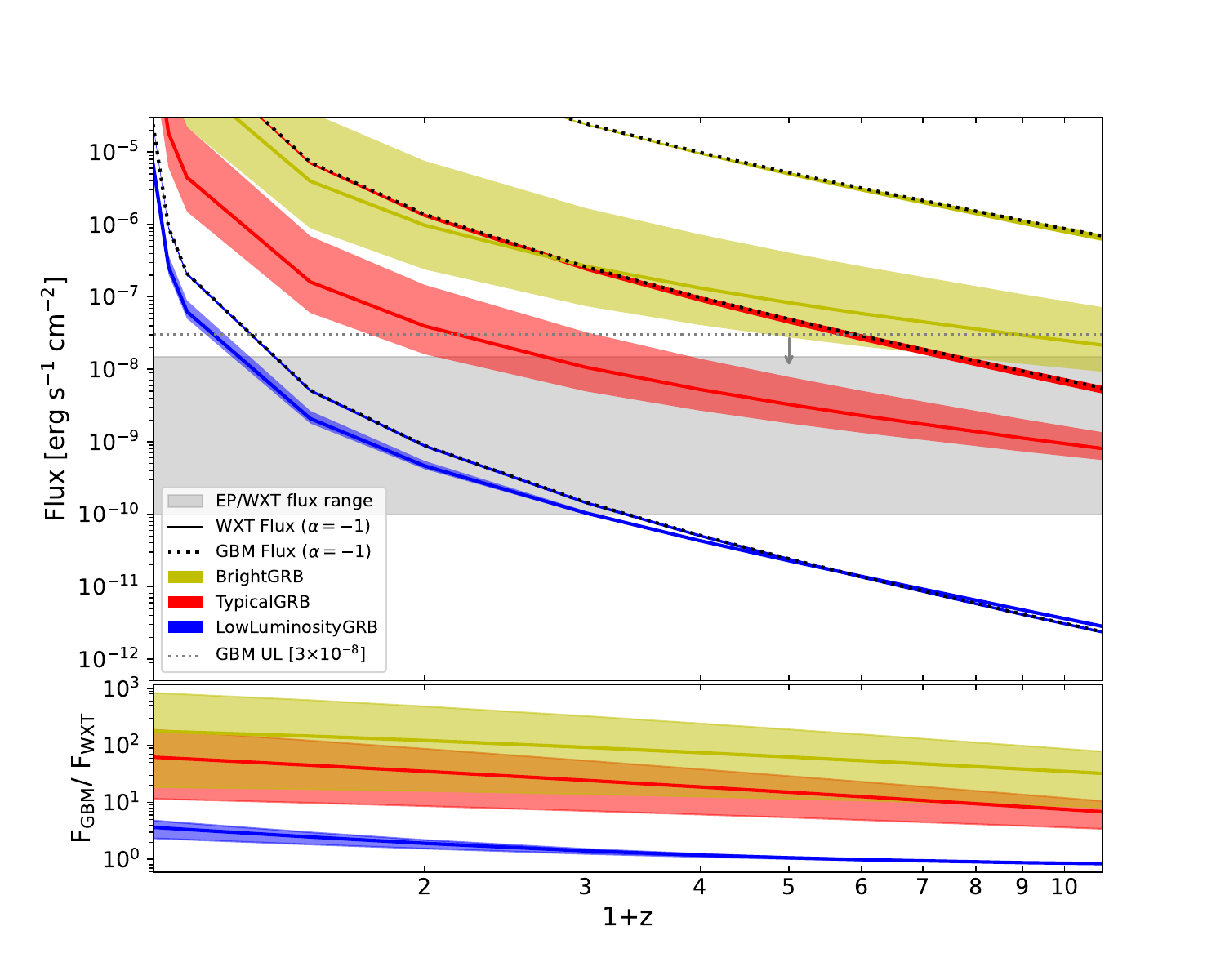}
   \caption{Expected fluxes in the GBM 10--1000 keV band and EP/WXT 0.5--4 keV band (top), and their ratio (bottom), as a function of redshift for three representative GRBs following the Yonetoku relation \citep{2004ApJ...609..935Y}, modelled with a Band spectrum (see Sec. $\S$\ref{sec:discussion_SED}). For each of the three cases, the solid curves correspond to $\alpha=-1$, while the shaded regions span $-1.5 \leq \alpha \leq -0.67$ for a fixed $\beta=-2.25$. The grey band shows the range of observed WXT average fluxes, and the dotted horizontal line indicates a representative average \textit{Fermi}/GBM flux upper limit of 3$\times10^{-8}$ erg~$\rm s^{-1}cm^{-2}$ obtained for the EP/WXT events.}

         \label{fig:band}
\end{figure}

In Section \ref{sec:GBM_results}, we show that the ratio between the simultaneous $\gamma$-ray and X-ray average fluxes typically lies in the range $1 \lesssim F^{\mathrm{avg}}_{\gamma}/F^{\mathrm{avg}}_{\mathrm{X-ray}} \lesssim 100$. 
No statistically significant correlation is found between the simultaneous X-ray and $\gamma$-ray fluxes  (Spearman $\rho = -0.25$, $p = 0.32$; Pearson $r = -0.04$, $p = 0.89$, for 18 events). This is not unexpected: if the two emission components share a common physical origin but sources span a range of spectral shape (in terms of peak energy, low-energy photon index, and luminosity), a correlation would still be expected, but with substantial intrinsic scatter.
Disentangling this from a scenario where X-ray and $\gamma$-ray emissions arise from physically distinct processes requires a detailed joint fit of both dataset, ideally over multiple time bins to follow the evolution of the different components, which is beyond the scope of this paper.
However, with the current sample of fluxes it is already possible to perform a simple consistency check between the observed flux ratio $F_{\gamma}$/$F_{X}$ and the predicted ones based on different spectral shapes. 
To this aim, we compute the expected fluxes in the GBM band [10-1000 keV] and in the WXT band [0.5-4 keV] assuming a Band spectrum with fixed high-energy photon index $\beta=-2.3$, as a function of redshift. Differently from the test described above (see Figure \ref{fig:fluxratios}), in this case we also included the known Yonetoku relation \citep{2004ApJ...609..935Y} between peak isotropic luminosity and peak energy. Three representative GRB cases are considered from the Yonetoku relation: a bright GRB with $L_{\rm iso,bol}=10^{54}\,\mathrm{erg\,s^{-1}}$ and $E_{\rm p,z}=1000$ keV, a typical GRB with $L_{\rm iso,bol}=10^{52}\,\mathrm{erg\,s^{-1}}$ and $E_{\rm p,z}=200$ keV, and a low-luminosity GRB with $L_{\rm iso,bol}=10^{49}\,\mathrm{erg\,s^{-1}}$ and $E_{\rm p,z}=10$ keV. We show the results in Figure~\ref{fig:band}. The solid curves correspond to $\alpha=-1$, while the shaded bands show the range obtained by varying the low-energy photon index between $\alpha=-0.67$ and $\alpha=-1.5$. The grey shaded region marks the range of average WXT fluxes measured for the EP sample, while the horizontal dotted lines indicate a representative threshold for GBM upper limits ($3 \times 10^{-8} \,\, \rm erg/cm^2/s$).

Figure~\ref{fig:band} shows that the predicted flux ratio $F_{\gamma}/F_{\rm X}$ varies significantly across the three GRB cases and evolves with redshift, naturally accounting for the scatter observed in Figure~\ref{fig:cfr_GBMavgflux}.
Also, it shows that flux ratios $F_{\gamma}/F_{X}\lesssim10$ are mainly reproduced by the low-luminosity GRB scenario, characterized by intrinsically low peak energies, and partially also by high-redshift typical GRBs. 
This result supports the interpretation proposed in the previous section using the GBM upper limits, where the low values of $F_{\gamma}/F_{X}$ point to spectra with low $E_{\rm peak}$. While that argument was based purely on the observed flux ratios and did not assume any intrinsic luminosity, here we incorporate the empirical correlation linking spectral peak energy and luminosity in the GRB population. If the Yonetoku relation holds also for lower luminosity events, this analysis strengthens the interpretation that at least a fraction of EP transients without detected high-energy emission may belong to the lower luminosity, lower $E_{\rm peak}$ population of GRBs.
An alternative, physically distinct scenario would invoke shock breakout emission, which is also expected to produce soft, $\gamma$-ray-faint spectra. We note that \citet{Fryer2026}, using a semi-analytic thermal Bremsstrahlung shock-breakout model, predict a ratio of energy emitted in the 10--1000\,keV and 1--10\,keV bands of $\sim0.1-6$ for Ic-BL supernovae shock-breakout (see their Table 4). Although this is a total energy output ratio over somewhat different energy bands and time windows than our averaged flux ratio $F_{\gamma}/F_{\rm X}$, it overlaps with events with flux ratios $F_{\gamma}/F_{X}\lesssim10$, suggesting shock-breakout emission from Ic-BL supernovae as a plausible contributor to this low-ratio population.

Moreover, the plot shows that, for the assumed spectral shape and Yonetoku scaling, an intrinsically bright GRB can be detected by WXT while remaining below the GBM upper limits only if placed at a very high redshift, typically $z\gtrsim 10$. Despite being detectable by WXT, no bright GRB has been detected by EP in its first year of activity, as they are rarer than the typical to lower luminosity ones \citep{PalmerioDaigne2021,GhirlandaSalvaterra2022}. 
In contrast, a typical long GRB event can satisfy both the observed WXT fluxes and the GBM non-detections only when placed at moderately high redshifts, $z\gtrsim 5$. Finally, a low-luminosity GRB naturally produces a much lower GBM-to-WXT flux ratio and can therefore account for the observed X-ray fluxes without violating the GBM constraints even at relatively nearby distances, $z\lesssim 0.3$. This shows that the absence of a GBM counterpart does not by itself exclude a GRB-like origin for the EP/WXT transients, but it favors either softer and less luminous events, moderately high-redshift typical GRBs, or very high-redshift bright GRBs.

\subsection{Comparison with the sample of Fermi/GBM GRBs} \label{sec:discussion_comparisonFermiGRB}

\begin{figure}
   \centering   \includegraphics[width=0.546\textwidth]{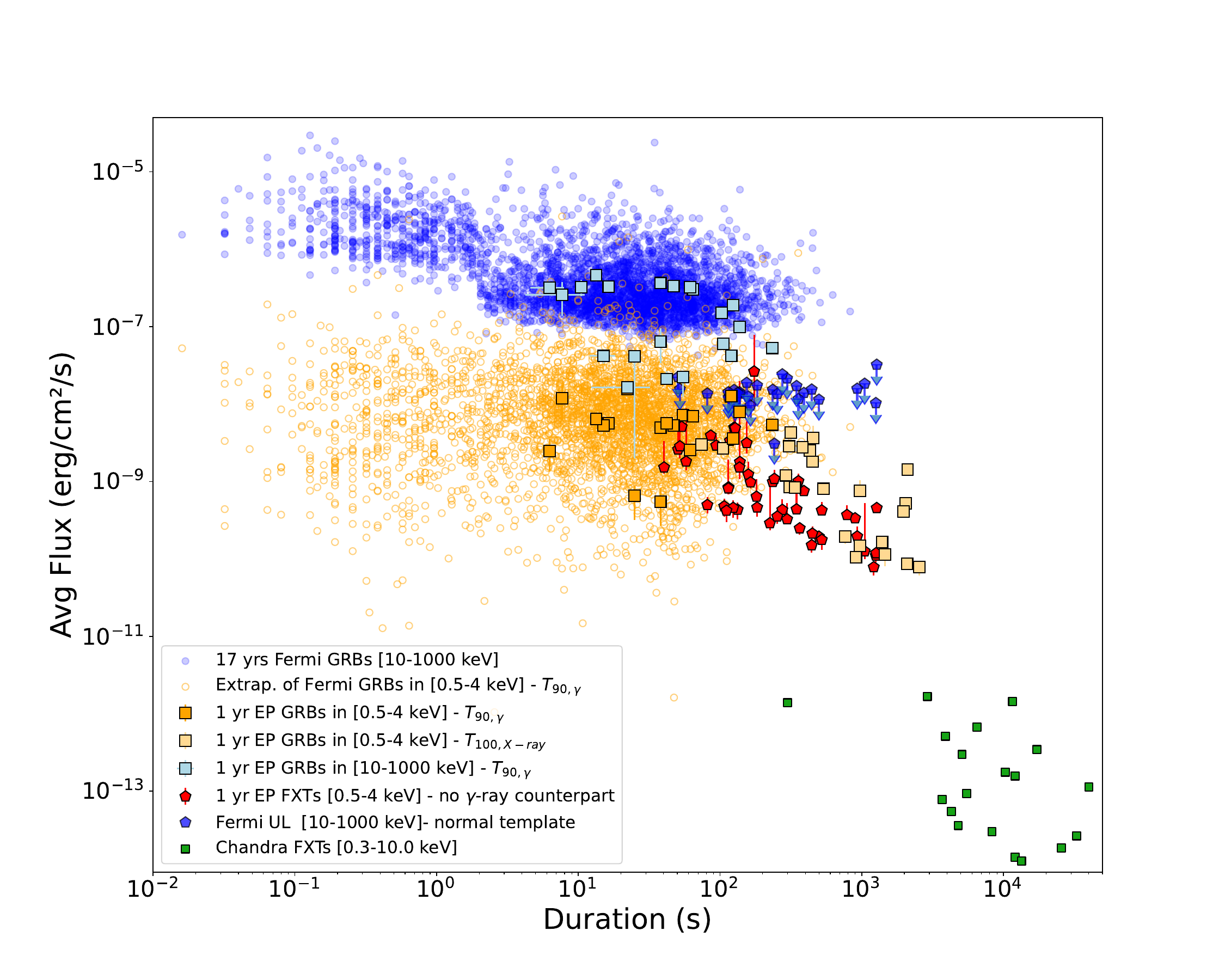}
      \caption{Time-averaged fluxes 
      versus durations of Fermi/GBM GRBs, EP/WXT sources, and Chandra FXTs. The average fluxes and $T_{\rm 90}$ of the entire sample of Fermi/GBM GRBs (3983 events) are shown with blue points. Their predicted counterparts in the WXT energy band [0.5-4 keV], obtained by extrapolating the best-fitting model from the Fermi/GBM Catalogue, are shown as orange open points. 
      EP sources with a $\gamma$-ray counterpart are plotted with orange squares if the X-ray flux is integrated over the $\gamma$-ray duration $T_{90, \gamma}$, while the beige squares represent their position when the X-ray flux is integrated over the total duration in X-ray $T_{100, \, X-ray}$. The light-blue squares represent their corresponding $\gamma$-ray counterparts. 
      EP sources without a $\gamma$-ray counterpart are plotted with red pentagons, and their corresponding upper limits in Fermi/GBM (using the \textit{normal} spectral template) are represented with blue pentagons and downward arrows. Note that there is no one-to-one correspondence, as not all EP events were visible by Fermi/GBM. 
      For comparison, Chandra-detected FXTs are shown with green points (taken from \citealt{2023QuirolaVasquez}).}
         \label{fig:extension_GBMcatalog}
\end{figure}

Figure~\ref{fig:extension_GBMcatalog} shows the distribution of time-averaged fluxes (computed over the total duration of the burst) as a function of the event duration for the classes of high-energy transients relevant to this work: \textit{Fermi}/GBM GRBs, EP/WXT sources, and \textit{Chandra} FXTs. The population of \textit{Fermi}/GBM GRBs, comprising $\sim$4000 events collected over 17 years of observations, is shown as blue points. 
EP sources with an identified $\gamma$-ray counterpart are highlighted with square markers. As described in Sec.~\ref{sec:GBM_results}, two different measurements of the X-ray flux are reported: orange squares correspond to X-ray fluxes integrated over the $\gamma$-ray duration ($T_{90, \gamma}$), while light beige squares represent fluxes computed over the total X-ray duration ($T_{100, \rm X-ray}$). Their associated $\gamma$-ray counterparts are shown as light-blue squares. 
EP sources without a $\gamma$-ray counterpart are instead marked with red pentagons, and, for those visible by GBM, the corresponding 3$\sigma$ upper limits derived from \textit{Fermi}/GBM data are indicated by blue pentagons with downward arrows (showing only the normal spectral template results for clarity purposes). We did not include the upper limit derived by \citealt{Zhang2025_upperlimits}, as in this plot they would be less informative than the GBM ones.  
For comparison, we also added the population of fast X-ray transients discovered in \textit{Chandra} via archival searches \citep{2022QuirolaVasquez,2023QuirolaVasquez}, shown as green points.

Some of the detected $\gamma$-ray counterparts (i.e. the light blue squares) extend the \textit{Fermi}/GBM GRB population towards the faint end, bridging the gap between detected bursts and the upper-limit regime. These events were identified through the Targeted Search, highlighting the role of EP in uncovering a previously hidden population of sub-threshold GRBs that remains largely unexplored. We note that the extension towards the faint and long end of the GRB population is, at least in part, a consequence of the different observational strategies of the two instruments: GBM's onboard trigger algorithm is limited to short timescales, and is therefore intrinsically less sensitive to longer-duration and low peak-flux events, while WXT's imaging capability allows it to accumulate signal over longer integration times, providing higher sensitivity to fainter, longer-duration transients.

To place the EP GRB sample in the context of the known GRB population, we investigated how the \textit{Fermi} GRB population would appear in the WXT energy band (0.5-4 keV). 
We extrapolated the best-fitting prompt-emission spectral models from the GBM catalogue (restricting to peaked functions, i.e. SBPL, Band, or CPL) into the X-ray regime, assuming no additional low-energy spectral breaks and no correction for X-ray absorption, consistent with the unabsorbed fluxes reported for the EP sample.
Since the extrapolated X-ray emission is assumed to arise from the same prompt spectral component, it is assigned the same duration $T_{\rm 90,\gamma}$. 
The resulting extrapolated population is shown as orange open circles in Figure~\ref{fig:extension_GBMcatalog}.
Comparing them with the simultaneous X-ray fluxes for EP GRB-associated events (orange squares), they occupy a remarkably similar region of the parameter space. A two-sample KS test yields $p_{\rm KS}=0.20$, suggesting that, despite the small size of the sample (18 events), the X-ray emission detected by EP is consistent with being the soft extrapolation of the same GRB average spectrum observed in $\gamma$-rays by Fermi. We caution that this test relies on the best-fit spectral models from the \textit{Fermi}/GBM catalogue, which may not always provide the best description of the observed spectra, and that the X-ray emission of individual sources could deviate from a simple extrapolation of the $\gamma$-ray spectral shape.\\

As already noted in Sect.~\ref{sec:GBM_results}, the distributions of time-averaged X-ray 
fluxes for EP sources with and without detected $\gamma$-ray counterparts (light beige squares vs red pentagons) are remarkably similar, when considering the full X-ray duration. 
Throughout this discussion, we have implicitly compared non-GRB EP events with the GRB population, but it is worth stressing that the physical nature of these sources remains unknown. Their X-ray emission could arise from a variety of physically distinct scenarios, including tidal disruption events, shock breakouts (relativistic and non), and cocoon emission. 
 
However, if such alternative populations are present in our sample of EP events without a $\gamma$-ray detection, which is entirely possible, it would be challenging to distinguish them from GRB-associated events, both in terms of X-ray flux and duration. Within the limitations of the present sample, no sub-population emerges as clearly distinct from the GRB-associated 
events in this parameter space. This may suggest that alternative origins can only constitute a minor contribution to the overall EP extragalactic transient population, which would instead be dominated by the population of spectrally softer GRBs. 
\\

Finally, it is interesting to compare the EP sources also with the population of FXTs discovered in \textit{Chandra} via archival searches \citep{2022QuirolaVasquez, 2023QuirolaVasquez}, shown as green points. As shown in Figure \ref{fig:extension_GBMcatalog}, \textit{Chandra} FXTs are 1-2 orders of magnitude longer in duration and 2-3 orders of magnitude fainter than EP transients, occupying the slow and faint corner of this parameter space. This raises the intriguing question of whether these populations are physically connected: if so, the observed flux-duration parameter space would be filled by a continuous sequence from fast, bright $\gamma$-ray transients, through the EP population in the intermediate regime, down to slow, faint X-ray transients. On the other hand, the apparent gap may also be influenced by observational selection effects. The much smaller field of view of \textit{Chandra} makes serendipitous detections of EP-like transients less likely, and previous archival searches were primarily optimised for longer-duration FXTs. EP is rapidly filling the previously unexplored bright and short-timescale region of this parameter space, but remains limited by its sensitivity. Bridging the gap from the other direction, namely searching for Chandra FXTs that are brighter and shorter than those found so far, may therefore be the key to address this open question.

\subsection{Constraints on energetics and $\Gamma_0$}\label{sec:discussion_Gamma0}

For each EP source with a detected $\gamma$-ray counterpart and with a measured redshift, we used the prompt $\gamma$-ray fluxes in the 10--1000 keV energy range to compute the pseudo-bolometric (1--$10^4$ keV rest-frame energy range), $k$-corrected isotropic luminosities $L_{\rm iso}$ and energies $E_{\rm iso}$, using the corresponding durations $T_{90,\gamma}$. 
For each EP source without a detected $\gamma$-ray counterpart and with a measured redshift, we used the flux upper limits derived using the normal spectral template. For the corresponding isotropic energies $E_{\rm iso}$, we used the observed duration of the X-ray emission $T_{\rm 100, X-ray}$.

While for the GRB-associated events, the derived $L_{\rm iso}$ and $E_{\rm iso}$ span approximately $L_{\rm iso}\sim7\times10^{50}-1\times10^{53}$ erg s$^{-1}$ and $E_{\rm iso}\sim6\times10^{51}-6\times10^{53}$ erg, the derived upper limits for those without $\gamma$-ray counterpart are in the range $L_{\rm iso}\lesssim1\times10^{48}-9\times10^{51}$ erg s$^{-1}$ and $E_{\rm iso}\lesssim5\times10^{49}-6\times10^{53}$ erg. While the upper limits on $E_{\rm iso}$, which takes into account the long duration of the X-ray transients, are not particularly constraining, those on $L_{\rm iso}$ mostly fall below the region populated by classical GRBs, with only a partial overlap at the faint end of the distribution.
The resulting values of $L_{\rm iso}$ and $E_{\rm iso}$ as a function of redshift, for events with and without a $\gamma$-ray counterpart, are shown in Figure~\ref{fig:EisoLiso_vs_z}.

Albeit with small-number statistics, also the redshift distributions of the two populations show a marked difference: median $z = 3.55$ for GRB-associated events (n=9) versus $z = 0.8$ for EP-only events (n=12), with both a KS test ($p = 0.012$) and an Anderson-Darling test ($p = 0.004$) supporting this difference. 
Notably, the majority of EP-only events (67\%) lie at redshifts below that of the lowest-redshift GRB-associated event ($z = 1.673$), pointing to a low-redshift tail in the population lacking high-energy counterparts (see Figure~\ref{fig:cumulative_z}).
One possible explanation for this low-redshift tail is that it traces an underlying low-luminosity population, preferentially detected only at low redshift since these events are too faint to be detected at larger distances. Within the EP-only sample, the low-redshift subset (below the lowest GRB redshift) has a median X-ray luminosity roughly two orders of magnitude lower than the higher-redshift EP-only events ($\sim 3\times10^{47}$ vs. $\sim 3\times10^{49}$ erg/s), consistent with this scenario. This is somewhat in tension with the conclusion of \citealt{OConnor2025}, who find no statistically significant difference between the redshift distributions of GRB-associated and non-GRB EP transients. Future studies with larger samples, accounting for gamma-ray coverage, might help establish whether this low-redshift tail is a genuine feature of the population.
\\  

\begin{figure}
   \centering
   \includegraphics[width=\hsize]{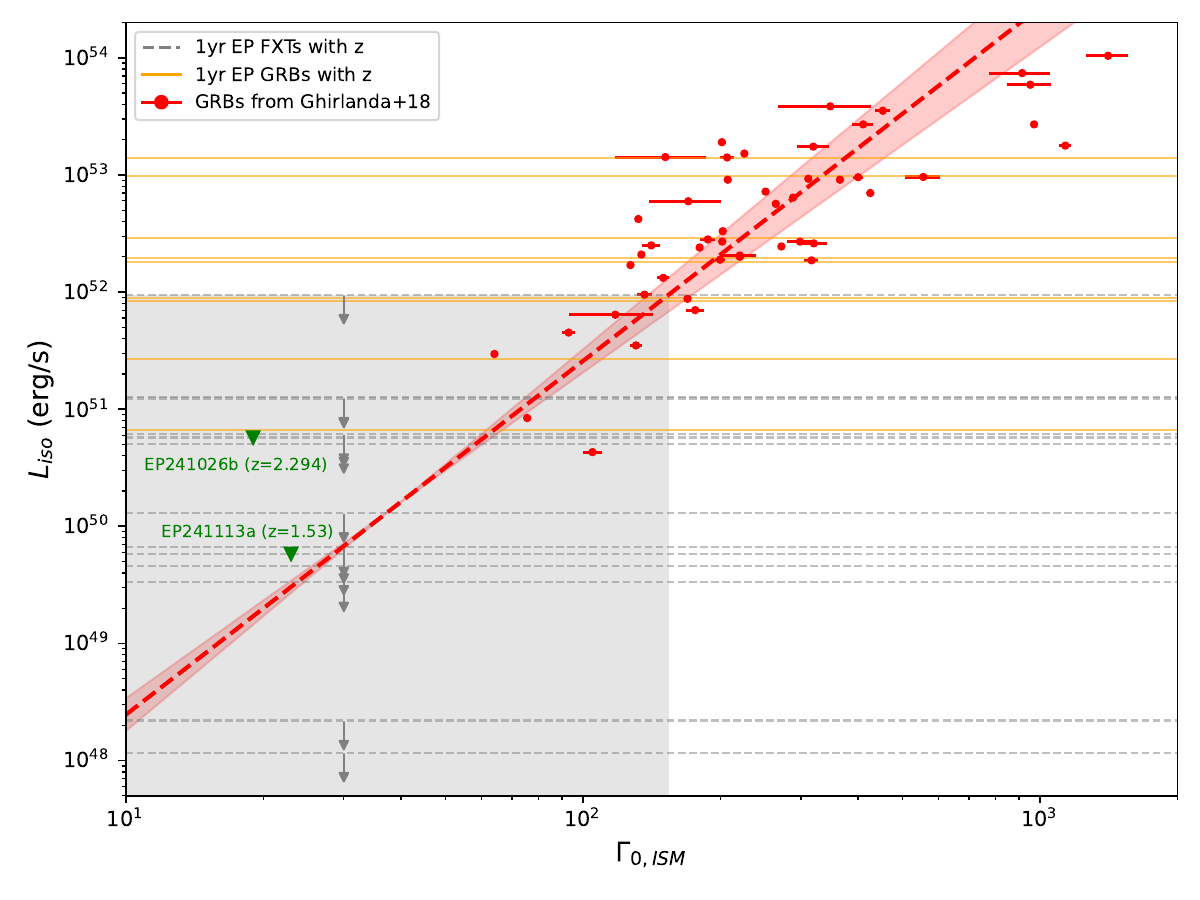}
      \caption{Comparison in the $\Gamma_0$--$L_{\rm iso}$ plane between the EP sources analyzed in this work and the GRB sample with well-constrained prompt emission properties and measured afterglow onset times from the 'gold' sample of \citealt{Ghirlanda2018} (red points). The dashed red line shows their reported best-fit $\Gamma_0$--$L_{\rm iso}$ correlation, with the shaded region indicating the 1$\sigma$ scatter. For the EP sources with and without $\gamma$-ray counterpart, orange solid lines indicate $L_{\rm iso}$ measurements, while grey dashed lines represent upper limits, respectively. Two EP events, EP241026b    \citep{Joyce_26b} and EP241113a \citep{Dai2026}, for which $\Gamma_0$ was estimated through afterglow modelling, are highlighted with green downward triangles, together with their redshift value.
      }\label{fig:GG18_correlations}
\end{figure}

If EP FXTs without detected $\gamma$-ray emission indeed correspond to softer and intrinsically less energetic events, their jet dynamics may also differ from those of classical GRBs, motivating an exploration of the constraints on their bulk Lorentz factors. Following the empirical $\Gamma_0$--$L_{\rm iso}$ 
correlation reported in \citet{Ghirlanda2018}, derived from a 'gold' sample of GRBs with well-constrained spectral parameters and $\Gamma_0$ estimates from afterglow deceleration peak times (red points and dashed line in Figure~\ref{fig:GG18_correlations}), the upper limits on $L_{\rm iso}$ derived for our sample can translate into upper limits on $\Gamma_0$.
A similar plot with $E_{\rm iso}$ is reported in Figure \ref{fig:GG18_correlations_appendix}.
If the same empirical correlations established for classical GRBs holds also to this regime, the derived upper limits on $L_{\rm iso}$ would correspond to lower initial Lorentz factors $\Gamma_0 \lesssim 100$ for this population. Probing this low-$\Gamma_0$ - low-$L_{\rm iso}$ corner of the parameter space is physically important: it allows us to test whether relativistic jets (both collapsar- or merger-driven events) form a continuous distribution in both bulk velocity and energy release and down to which limit. 

So far, this region has remained largely unexplored, as classical $\gamma$-ray triggering instruments are insensitive to the faint or soft prompt emission expected from such events. The sensitivity of EP to soft X-ray transients is already delivering a handful of events with direct constraints on $L_{\rm iso}$ and $\Gamma_0$ in this regime, marked with green downward triangles in Figure~\ref{fig:GG18_correlations}.
A first example is EP241026b, which lacks a $\gamma$-ray counterpart ($L_{\rm iso}<5.7 \times 10^{50} \,\, \rm erg/s$). A detailed broadband modelling of the afterglow favoured an on-axis jet with a $\Gamma_0 \sim$ 19 (\citealt{Joyce_26b}).
This might also be the case for another example, EP241113a, whose afterglow modelling implies an on-axis jet with $\Gamma_0 \sim$23 \citep{Dai2026}, and no $\gamma$-ray detection despite an isotropic energy (calculated only from X-rays) of $E_{\rm iso} \sim 1.4 \times 10^{51}$ erg. A more complex case is represented by EP240414a, a fast X-ray transient at $z=0.401$ associated with a Type Ic-BL supernova \citep{vanDalen2025, Sun2025, Bright2025, Srivastav2025}, which has been modelled in the literature under two physically distinct scenarios. In the off-axis interpretation, multiwavelength afterglow fitting yields $\Gamma_0 \sim$ 61 \citep{Ren2026}, consistent with a standard relativistic jet whose $\gamma$-ray emission is fainter due to unfavourable viewing geometry. In the on-axis jet interpretation, in which the jet propagates through extended circumstellar material and undergoes significant baryon loading, the inferred Lorentz factor is instead $\Gamma_0 \sim$20 \citep{Hamidani2025}. 
A similar case is represented by EP241021a ($L_{\rm iso}<4.6 \times 10^{49} \,\, \rm erg/s$), suggested to be associated with a collapsar \citet{Quirola-Vasquez2026}, which has been modelled with a mildly relativistic ($\Gamma_0 \sim 40$) jet, or alternatively, a highly relativistic ($\Gamma_0 \gtrsim 100$) structured jet seen substantially off-axis \citep{Yadav2025, Gianfagna2025}.
Distinguishing between these two scenarios has direct implications for our understanding of jet 
structure and would inform GRB population models. Independent constraints on the viewing angle, e.g. from very long baseline interferometry (VLBI) measurements of apparent source motion or from compactness arguments or from a GW signal, are essential to break this degeneracy.
Together, these events (mostly with SN detections) are 
already providing the first opportunities to probe the mildly relativistic regimes and test jet models in a regime of energetics that was previously inaccessible (see also \citealt{Chen2026_lowGamma}).

\subsection{Implications on the prompt SSC component}\label{sec:discussion_SSC}

The low spectral peak energies and the low initial Lorentz factors inferred for EP sources without $\gamma$-ray counterparts in Section \ref{sec:discussion_Gamma0} also have a direct implication for potential high-energy emission. Assuming GRBs as astrophysical sources powered by synchrotron radiation, in the standard synchrotron plus synchrotron self-Compton (SSC) scenario, the SSC component peaks at an energy $E_{\rm SSC, peak} \propto \gamma_m^2 E_{\rm sync, peak}$, where $E_{\rm sync, peak}$ is the synchrotron peak energy \citep{RybickiLightman1979}. 
While for typical long GRBs the SSC component is expected to peak in the TeV band (see \citealt{Nava2021HEreview, Miceli2022} for recent reviews), placing it beyond the reach of space-based $\gamma$-ray instruments and making the detection of the prompt possible only for ground-based Cherenkov telescopes, for less energetic events the combination of lower Lorentz factors and lower synchrotron peak energies both act to shift the SSC component to significantly lower energies, making it more dominant in the GeV band, where it becomes directly testable with \textit{Fermi}/LAT. A typical GRB would emit much more luminosity in the synchrotron and SSC components, but most of its SSC would fall in the TeV energy range; also, a typical GRB is more frequently found at a higher redshift than low-luminosity events, implying also a stronger suppression due to the extragalactic background light (EBL) opacity. 
Motivated by this, we investigate whether the LAT upper limits obtained for our sample can help in constraining the presence of a second high-energy spectral component produced through SSC emission.

To explore this scenario, we use the lepto-hadronic radiative code \texttt{LeHaMoC} \citep{Stathopoulos2024}, adopting parameters representative of a low-luminosity GRB, with $L_{\rm iso}=10^{49}\,\mathrm{erg\,s^{-1}}$ and rest frame peak energy $E_{\rm p,z}=10$ keV located at $z=0.2$. The recent identification of the synchrotron cooling break frequency in the hard X-ray/$\gamma$-ray regime \citep{Oganesyan2018, Ravasio2019} implied that the required magnetic field must be much smaller ($B^\prime \sim 1$--$10\,\mathrm{G}$) than what is usually assumed ($B^\prime \sim 10^4$--$10^6\,\mathrm{G}$). Therefore, in our test we explored different values of the comoving magnetic field spanning from 1 G to $10^4$ G. The variability timescale in the observer frame was assumed to be $t_{\rm var}=1$ s, corresponding to an emission radius $R$ of a few $\times$10$^{13}$ - $\times$10$^{16}$ cm, depending on $\Gamma$. We additionally adopted a steep injected electron distribution with power-law index $p=3$. Such a choice reduces the relative contribution of the highest-energy electrons, thereby mitigating the importance of Klein–Nishina effects. 
We then computed the observable fluxes in the LAT and WXT energy ranges, for each combination of bulk Lorentz factor and comoving magnetic field.

\begin{figure*}
    \centering
    \begin{subfigure}[b]{0.49\textwidth}
        \includegraphics[width=\textwidth]{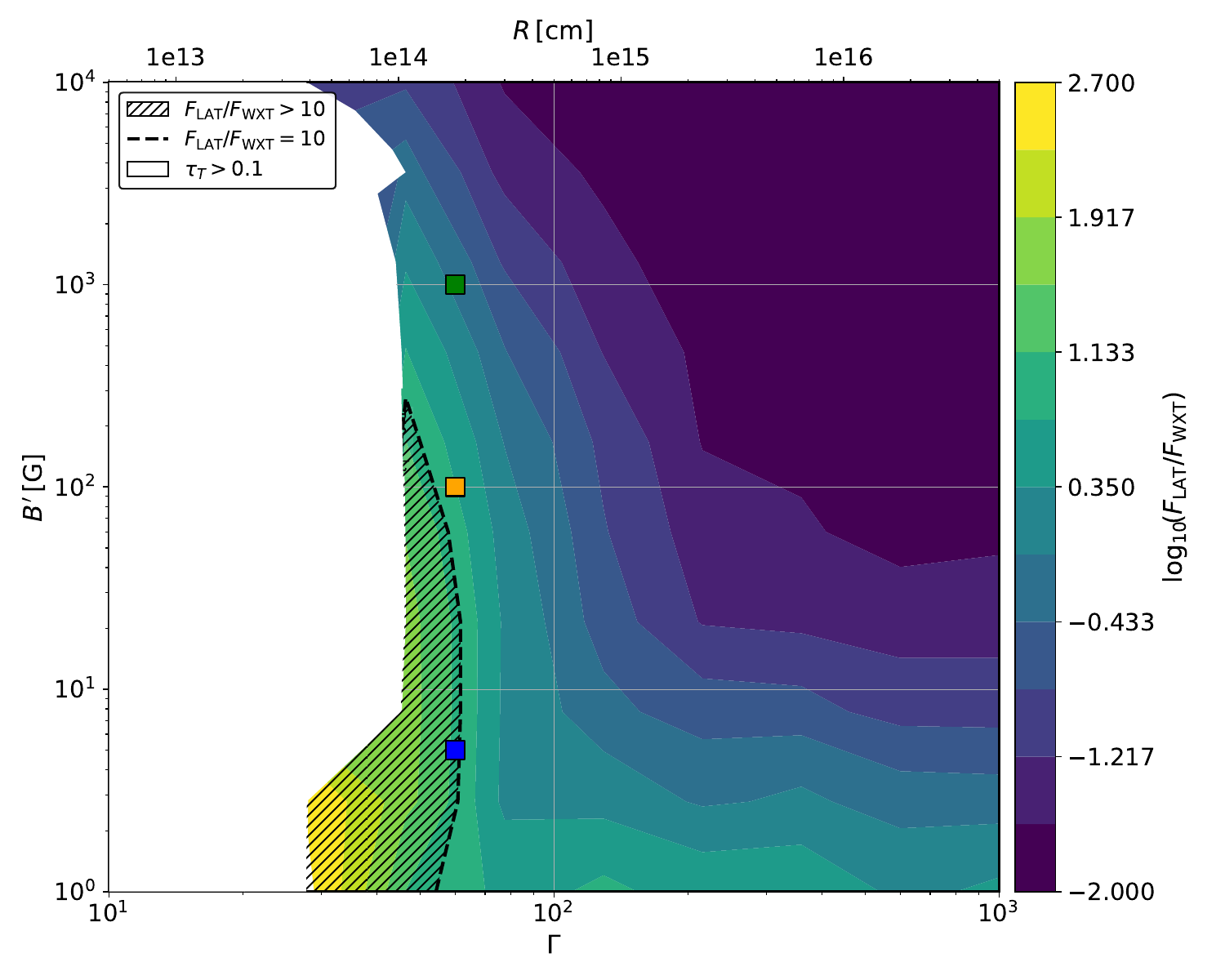}
        \caption{}
        \label{fig:ssc_2b}
    \end{subfigure}\hfill
    \begin{subfigure}[b]{0.49\textwidth}
        \includegraphics[width=\textwidth]{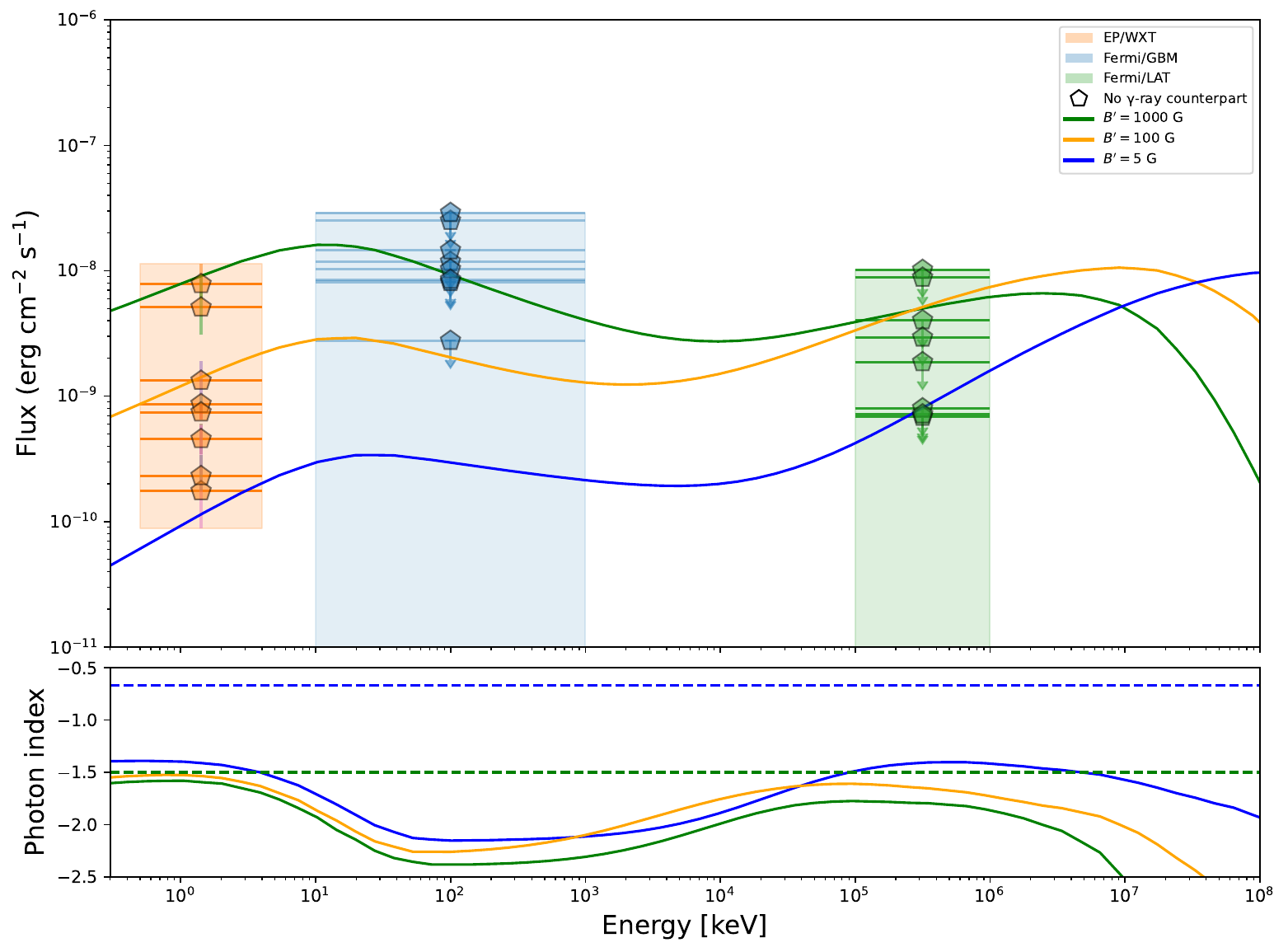}
        \caption{}
        \label{fig:ssc_2a}
    \end{subfigure}\hfill
    \caption{SSC modelling of a low-luminosity GRB-like scenario. 
\textit{Left}: $B'-\Gamma$ parameter space colour-coded by the predicted flux ratio $F_{0.1-1\,{\rm GeV}}/F_{0.5-4\,{\rm keV}}$, assuming a low-luminosity GRB with $L_{\rm iso}=10^{49}\,\mathrm{erg\,s^{-1}}$ and rest frame peak energy $E_{\rm p,z}=10$ keV, located at $z=0.2$, with a variability timescale of $t_{\rm var,obs}=1$ s. The secondary horizontal axis shows the corresponding emission radius, $R=2c\Gamma^2 t_{\rm var,obs}/(1+z)$. The black dashed contour marks a flux ratio of 10, and the shaded region indicates where the predicted fluxes in LAT exceeds ten times the WXT band fluxes, which is disfavored by the LAT upper limits derived in this work. The white region corresponds to parameter space with $\tau_{\rm T}>0.1$, where the source becomes optically thick to Thomson scattering and the spectra are not considered reliable in the present calculation. The square markers indicate the three parameter choices shown in the left panel.
\textit{Right}: Model spectra computed with \texttt{LeMoC} for a source at $z=0.2$, adopting $\Gamma=60$, $t_{\rm var,obs}=1$ s, and three values of the comoving magnetic field, $B'=5$, 100, and 1000 G. The spectra are compared with the observed WXT fluxes, GBM fluxes or upper limits, and LAT upper limits for the EP/WXT events within the LAT field of view. Bottom panel shows the corresponding photon index for each spectra.
}
    \label{fig:SSC_2}
\end{figure*}

Figure~\ref{fig:ssc_2b} shows the $B'-\Gamma$ parameter space, colour-coded by the ratio between the predicted LAT-band and WXT-band fluxes, $F_{0.1-1\,{\rm GeV}}/F_{0.5-4\,{\rm keV}}$. The top horizontal axis gives the corresponding emission radius, computed as $R=2c\Gamma^2 t_{\rm var,obs}/(1+z)$ for the same variability timescale and redshift described above.
We reject the parameter space where the emission region is optically thick to Thomson scattering, $\tau_{\rm T}>0.1$; in this regime, non-linear radiative effects become important, and the present LeMoC setup does not provide reliable spectra. The black dashed contour marks $F_{0.1-1\,{\rm GeV}}/F_{0.5-4\,{\rm keV}}=10$, while the shaded region indicates where this ratio exceeds 10. Since the LAT upper limits are generally below ten times the measured WXT fluxes (Fig.~\ref{fig:cfr_LATavgflux}), this region is excluded by the LAT non-detections. It corresponds mainly to lower values of $\Gamma$ and $B'$, where the SSC component becomes comparatively stronger. Thus, in the scenario of a low-luminosity GRB, the LAT upper limits derived for some of the EP sources provide an additional constraint on the parameter space, disfavouring combinations of low $B'$ and low $\Gamma$ within the adopted model assumptions.

As illustrative examples, we selected three representative cases with $B'=5$, 100, and 1000 G, adopting a fixed bulk Lorentz factor of $\Gamma=60$. We then computed the corresponding spectra and compared them with the observed SEDs derived in Sec. \ref{sec:SED_results} (see also Figure \ref{fig:SED}). The resulting spectra are shown in Fig.~\ref{fig:ssc_2a}, together with the observed WXT fluxes, the GBM fluxes or upper limits, and the LAT upper limits, for the subset of EP/WXT events without a $\gamma$-ray counterpart that were visible by LAT (8 events). For the explored parameter space, the predicted SSC component remains below the LAT upper limits, indicating that the absence of detected GeV emission does not rule out the presence of an additional SSC spectral component associated with the EP/WXT transients.
In the near future, the Cherenkov Telescope Array Observatory (CTAO) is expected to have better sensitivity with an exposure of about 100 seconds in the few tens of GeV range (in the energy band around 25 GeV) than LAT by at least four orders of magnitude\footnote{See for example, \url{https://www.ctao.org/for-scientists/performance/.}}.

However, we emphasize that the analysis presented here should be regarded as an exploratory assessment rather than a comprehensive modeling effort. The physical conditions governing on-axis prompt GRB emission remain highly uncertain. Key quantities, such as the location of the emitting region, the comoving magnetic field strength, the bulk Lorentz factor, and the characteristic particle energies, are still poorly constrained.
As a result, the expected ratio between SSC and synchrotron luminosities remains difficult to predict. 
Moreover, high-energy emission in GRBs may arise from additional processes beyond SSC, including inverse Compton scattering of external seed photons and hadronic mechanisms, whose investigation is beyond the scope of this work.  
Therefore, a broader exploration of the parameter space, accounting for the large uncertainties in the physical conditions of the emitting region and alternative emission scenarios, will be necessary to fully evaluate the constraints imposed by the LAT upper limits derived in this study.

\section{Conclusions}\label{sec:conclusions}

We presented a systematic search for prompt $\gamma$-ray counterparts associated to the 72 transients detected by EP/WXT during its first year of operation (February 2024 -- February 2025), combining observations from \textit{Fermi}/GBM in the energy range 10-1000 keV and \textit{Fermi}/LAT in the energy range 0.1-1 GeV.

Our main findings can be summarised as follows:

\begin{enumerate}

\item As reported in Table \ref{tab:EP_stats}, 23 sources (32\% of the sample) are associated with a confirmed $\gamma$-ray counterpart, seven of which were identified sub-threshold (five via the Fermi Targeted Search used in this work, two tentative associations in Konus-Wind data) and would have been missed without the independent trigger form EP. 
Nine of the 23 GRB-associated events have a spectroscopic redshift (see Table \ref{tab:Gamma_counterparts}).
The majority of the EP sources (49 events, $68\%$ of the sample) remain undetected at $\gamma$-ray energies, despite most of them being visible by $\gamma$-ray instruments; 14 out of 49 non-GRB events have a spectroscopic redshift. For the events without a $\gamma$-ray counterpart, we derived $3\sigma$ flux upper limits in GBM (10--1000 keV - see Table \ref{tab:NoGamma_counterparts}) and in LAT (0.1-1 GeV - see Table \ref{Table_LAT}).
No significant GeV emission was detected by \textit{Fermi}/LAT for any of the 12 EP transients satisfying our LAT selection criteria.

\item EP transients without detected $\gamma$-ray counterparts exhibit time-averaged X-ray fluxes distributions statistically similar to those of GRB-associated events (KS test, $p_{\rm KS}=0.60$).  
This implies that the average X-ray brightness alone is not a reliable predictor of detectable $\gamma$-ray emission. The detectability in the $\gamma$-ray band is likely governed by additional parameters, in particular the spectral peak energy $E_{\rm peak}$ and the low-energy photon index.
Instead, in case of a GRB association, measuring the X-ray flux over the interval simultaneous with the $\gamma$-ray emission ($T_{90,\gamma}$) shifts it systematically to higher values, suggesting that the $\gamma$-ray emission preferentially occurs during episodes of enhanced X-ray activity. As shown in Figure \ref{fig:cfr_GBMavgflux} and Figure \ref{fig:cfr_LATavgflux}, the ratio between the simultaneous $\gamma$-ray and X-ray average fluxes lies in the range $1 \lesssim F_{\gamma}/F_{\rm X-ray} \lesssim 100,\,$ while GeV constraints indicate $1 \lesssim F_{\gamma,\rm LAT}/F_{WXT} \lesssim 10$. These ratios provide direct constraints on the broadband SEDs extending from X-rays to GeV energies, useful for identifying possible high-energy emission components;

\item The simultaneous X-ray fluxes of EP-GRBs, computed over the $\gamma$-ray duration, are statistically consistent with the X-ray fluxes predicted by extrapolating the best-fit prompt spectral models of \textit{Fermi}/GBM GRBs into the WXT energy band (KS test, $p_{\rm KS}=0.20$). This suggests that, for these events, the X-ray emission detected by EP is broadly consistent with the extension at low-energies of the Fermi GRB population (see Sect.\ref{sec:discussion_comparisonFermiGRB} and Figure \ref{fig:extension_GBMcatalog}). Several EP-GRBs identified via the Targeted Search represent the faint, sub-threshold extension of the classical \textit{Fermi}/GBM GRB population;

\item By comparing the measured X-ray fluxes with the GBM upper limits, we find that many EP events without $\gamma$-ray counterparts are consistent with softer prompt spectra characterised by relatively low peak energies. For the subset of non-GRB EP events with the most stringent $\gamma$-ray upper limits (i.e. where $F_{\gamma}/F_{\rm X-ray} \sim 1$--10), the GBM non-detections imply $E_{\rm peak} \lesssim 30$--100\,keV, depending on the assumed low-energy photon index of the Band function (see Fig.\ref{fig:fluxratios}). A comparison with the \textit{Yonetoku} relation indicates at least part of the EP population without detected $\gamma$-ray emission may belong to the low-luminosity, low-$E_{\rm peak}$ GRB regime (see Sect.\ref{sec:discussion_SED}).

\item For the 14 non-GRB EP events with a spectroscopic redshift, the derived upper limits on the isotropic-equivalent luminosity $L_{\rm iso}$ and energy $E_{\rm iso}$ in the 10--$10^4$\,keV band are $L_{\rm iso}\lesssim1\times10^{48}-9\times10^{51}$ erg s$^{-1}$ and $E_{\rm iso}\sim5\times10^{49}-6\times10^{53}$ erg, which supports their interpretation as low energetics events. If the empirical correlation $\Gamma_0$--$L_{\rm iso}$ found in \citet{Ghirlanda2018} holds in this regime, these results imply $\Gamma_0 \lesssim 100$ for this population. 
Several individual EP transients with independent afterglow-based estimates of $\Gamma_0$ already support this picture, indicating mildly relativistic outflows with $\Gamma_0\sim20$--60 (see Sect.\ref{sec:discussion_Gamma0}).

\item In Sect.\ref{sec:discussion_SSC},  we also investigated whether the derived LAT upper limits constrain the presence of an additional prompt SSC component. Under a representative modelling of a low-luminosity GRB case, the current LAT non-detections disfavour combinations of low comoving magnetic fields and low Lorentz factors, that would instead predict comparatively bright GeV emission.
The follow-up of EP sources with future Cherenkov Telescope Array (CTA) observations is crucial to extend these constraints and probe the GeV--TeV regime for this newly discovered population.

\end{enumerate}

The physical nature of individual EP transients without $\gamma$-ray counterparts remains uncertain: tidal disruption events, shock breakouts, cocoon emission, and other classes of X-ray transients are all viable alternatives.
However, under the hypothesis that these sources belong to the same physical population as GRBs, the multi-wavelength constraints derived in this work paint a consistent picture: the $\gamma$-ray upper limits favor intrinsically soft spectra with $E_{\rm peak} \lesssim 30$--100\,keV, the derived upper limits on $L_{\rm iso}$ and $E_{\rm iso}$ are consistent with the lower luminosity tail of the GRB energetics distribution, and the inferred bulk Lorentz factors $\Gamma_0 \lesssim 100$ place these events in the mildly relativistic regime. In other words, if these sources are GRBs, they would be consistent with the lower $E_{\rm peak}$, lower luminosity, less relativistic counterparts of the classical population, if seen on-axis. Alternatively, some may be classical GRBs observed off-axis, where the viewing geometry rather than intrinsic weakness suppresses the $\gamma$-ray emission. 
Discriminating between these scenarios is essential to understand both the structure of relativistic jets and the GRB luminosity function at low energies --- two key yet poorly constrained pieces of the GRB puzzle. 

As the EP sample continues to grow, statistical analyses of larger and homogeneous catalogues, combined with dedicated multi-wavelength follow-up, will be essential to discriminate between scenarios, constrain the $E_{\rm peak}$, $L_{\rm iso}$ and $\Gamma_0$ distributions, and ultimately determine what fraction of EP extragalactic transients truly belong to the GRB population, and what fraction may represent the emergence of an entirely new class of high-energy phenomena.
The necessary crucial next step, especially for the non-GRB associated events, is quantifying how the observed EP population maps onto theoretical expectations from the GRB population models. Assessing where GRB population models succeed in reproducing the data, and where they fail, will directly inform whether the hints inferred in this work from observations are corroborated (i.e. we are observing a continuous extension of the GRB population) or the necessity of additional physical channels. We will address this in detail through a dedicated comparison with GRB population synthesis models in a forthcoming publication.

\begin{acknowledgements}
      The authors would like to thank Dylan Adrien for kindly providing the gamma-ray properties of the SVOM-detected burst EP241113b. M.E.R. received the support of the Junior Leader Fellowship from ”la Caixa” Foundation (ID 100010434). The fellowship code is LCF/BQ/PI25/12100030. This work was also partly supported by the Spanish program Unidad de Excelencia María de Maeztu CEX2020-001058-M, financed by MCIN/AEI/10.13039/501100011033, and by the MaX-CSIC Excellence Award MaX4-SOMMA-ICE.
      M.E.R., P.G.J. J.N.D.D., J.S.S., J.Q.V., and A.P.C.H. are supported by the European Union (ERC, Starstruck, 101095973, PI Jonker). Views and opinions expressed are however those of the author(s) only and do not necessarily reflect those of the European Union or the European Research Council Executive Agency. Neither the European Union nor the granting authority can be held responsible for them. G.O. acknowledges support from the ASI-GSSI contract n. 2025-5-U.0: “Gamma-ray bursts: a probe of multi-messenger and extreme Astrophysics”. S.R. acknowledges support from the Astrophysics Center for Multi-messenger Studies in Europe (ACME), funded under the European Union’s Horizon Europe Research and Innovation Program, Grant Agreement No. 101131928. SVOM is a joint Chinese-French mission led by the China National Space Administration (CNSA),  the French Space Agency (CNES), and the Chinese Academy of Sciences (CAS). We gratefully acknowledge the support from NSSC, IAMCAS, XIOPM, NAOC, IHEP, CNES, CEA, and CNRS.
\end{acknowledgements}

\bibliographystyle{aa} 
\bibliography{references} 

\onecolumn
\begin{appendix} 
\section{Tables}

\begin{table}[H]
\centering
\tiny
\setlength{\tabcolsep}{3pt}
\renewcommand{\arraystretch}{1.0}

\begin{threeparttable}
\caption{List of FXTs detected by Einstein Probe from February 2024 to February 2025, with a $\gamma$-ray counterpart.}\label{tab:Gamma_counterparts}

\begin{tabular}{@{}lccccccccccc@{}}
\toprule
Name & Visible & $T_{100, \rm X-ray}$ & $F^{T_{100, X-ray}}_{\rm avg,X\text{-}ray}$ & $T_{90, \gamma}$ & $F^{\rm T_{90, \gamma}}_{\rm avg,X\text{-}ray}$ & $F_{\rm avg,\gamma-ray}$ & $F_{\rm peak,X\text{-}ray}$  & $F_{\rm peak,\gamma-ray}$ & z & $L_{\rm iso}$\\
 & by GBM? & [s] & [$10^{-9} \, \rm erg/cm^{2}/s$] & [s] & [$10^{-9} \, \rm erg/cm^{2}/s$] & [$10^{-7} \, \rm erg/cm^{2}/s$] & [$10^{-8} \, \rm erg/cm^{2}/s$] & [$10^{-7} \, \rm erg/cm^{2}/s$] & & [$10^{52} \, \rm erg/s$]\\
 \midrule
 \midrule
EP240202a\tnote{(0)} & $\times$ & 764.8 & $0.19_{-0.04}^{+0.03}$ & - & - & - & $0.28 \pm 0.21$ & - & - & -\\ \midrule
EP240219a & $\checkmark$ & 449.95 & $1.80_{-0.30}^{+0.63}$ & 54.8 & $7.15_{-0.77}^{+0.87}$ & $0.22_{-0.10}^{+0.01}$ & $3.02 \pm 1.43$ & $0.66_{-0.60}^{+0.14}$ & - & -\\ \midrule
EP240315a\tnote{(1)} & $\times$ & 2047.1 & $0.52_{-0.03}^{+0.05}$ & 38.0 & $4.95_{-0.79}^{+0.77}$ & $3.65_{-0.60}^{+0.75}$ & $2.69 \pm 1.44$ & $8.55_{-1.57}^{+1.69}$ & 4.859\tnote{(a)} & 9.87\\ \midrule
EP240617a & $\checkmark$ & 316.7 & $4.29_{-0.24}^{+0.31}$ & 120.0 & $12.70_{-0.74}^{+0.80}$ & $0.42_{-0.03}^{+0.03}$ & $3.20 \pm 1.43$ & $1.06_{-0.98}^{+0.01}$ & - \\ \midrule
EP240703a\tnote{(2)} & $\times$ & 453.3 & $3.64_{-0.61}^{+1.63}$ & 138.0 & $7.91_{-0.76}^{+0.74}$ & $0.99_{-0.13}^{+0.13}$ & $8.46 \pm 4.06$ & $2.80_{-0.40}^{+0.40}$ & - \\ \midrule
EP240703b\tnote{(0)} & $\times$ & 535.6 & $0.80_{-0.06}^{+0.09}$ & - & - & - & $0.66 \pm 0.48$ & - & - & -\\ \midrule
EP240801a & $\checkmark$ & 307.75 & $2.84_{-0.18}^{+0.22}$ & 22.3 & $15.70_{-1.56}^{+1.69}$ & $0.17_{-0.03}^{+0.03}$ & $4.20 \pm 1.43$ & $0.14_{-0.21}^{+0.21}$ & 1.673\tnote{(b)}  & 0.07\\ \midrule
EP240802a\tnote{(3)} & $\times$ & 429.8 & $2.51_{-0.18}^{+0.20}$ & 47.0 & $5.28_{-0.94}^{+1.01}$ & $3.34_{-0.15}^{+0.16}$ & $3.95 \pm 2.39$ & $82.70_{-26.30}^{+30.90}$ & - \\ \midrule
EP240804a\tnote{(4)} & $\times$ & 2089.75 & $0.09_{-0.01}^{+0.01}$ & 38.0 & $0.55_{-0.28}^{+0.28}$ & $0.64_{-0.31}^{+0.31}$ & $0.32 \pm 0.25$ & $2.06_{-0.90}^{+0.90}$ & 3.662\tnote{(c)} & 0.88 \\ \midrule
EP240807a & $\checkmark$ & 73.85 & $2.99_{-0.50}^{+2.47}$ & 7.68 & $11.90_{-5.71}^{+5.71}$ & $2.58_{-1.20}^{+1.20}$ & $1.79 \pm 1.08$ & $3.69_{-1.91}^{+1.91}$ & - \\ \midrule
EP240913a & $\checkmark$ & 973.25 & $0.15_{-0.03}^{+0.03}$ & 6.27 & $2.46_{-1.11}^{+1.09}$ & $3.16_{-0.35}^{+0.35}$ & $0.82 \pm 0.61$ & $9.85_{-0.39}^{+0.39}$ & - \\ \midrule
EP240919a & $\checkmark$ & 311.35 & $0.85_{-0.11}^{+0.11}$ & 25.0 & $0.65_{-0.34}^{+0.24}$ & $0.42_{-0.40}^{+0.03}$ & $0.26 \pm 0.16$ & $0.84_{-0.82}^{+0.08}$ & - \\ \midrule
EP240930a & $\checkmark$ & 382.05 & $2.76_{-0.24}^{+0.53}$ & 64.51 & $6.92_{-0.60}^{+0.67}$ & $3.04_{-0.33}^{+0.33}$ & $2.68 \pm 1.61$ & $14.50_{-2.53}^{+2.53}$ & - \\ \midrule
EP241025a & $\checkmark$ & 1974.2 & $0.41_{-0.02}^{+0.03}$ & 123.91 & $3.59_{-0.24}^{+0.25}$ & $1.89_{-0.14}^{+0.14}$ & $1.31 \pm 0.65$ & $3.34_{-0.83}^{+0.83}$ & 4.200\tnote{(d)} & 1.82\\ \midrule
EP241026a & $\checkmark$ & 291.75 & $1.20_{-0.12}^{+0.19}$ & 16.38 & $5.66_{-1.07}^{+1.09}$ & $3.27_{-0.22}^{+0.22}$ & $1.73 \pm 0.94$ & $16.00_{-5.85}^{+5.85}$ & 2.791\tnote{(e)} & 2.89 \\ \midrule
EP241113b\tnote{(5)} & $\times$ & 105.0 & $2.69_{-0.40}^{+0.96}$ & 42
& $5.61_{-1.22}^{+1.20}$ & $0.21^{+0.05}_{-0.03}$ & $1.98 \pm 1.44$ & $0.66^{+0.07}_{-0.07}$ & - \\ \midrule
EP241217b & $\checkmark$ & 2106.15 & $1.42_{-0.03}^{+0.05}$ & 234.0 & $5.41_{-0.24}^{+0.23}$ & $0.53_{-0.02}^{+0.02}$ & $2.47 \pm 0.87$ & $0.45_{-0.45}^{+0.23}$ & 1.879\tnote{(f)} & 0.27\\ \midrule
EP250109a & $\checkmark$ & 971.9 & $0.75_{-0.10}^{+0.28}$ & 15.1 & $5.27_{-1.43}^{+1.44}$ & $0.42_{-0.09}^{+0.09}$ & $1.86 \pm 1.35$ & $1.86_{-0.32}^{+0.32}$ & - \\ \midrule
EP250215a & $\checkmark$ & 1391.5 & $0.17_{-0.02}^{+0.03}$ & 13.37 & $6.42_{-1.30}^{+1.45}$ & $4.63_{-0.15}^{+0.15}$ & $2.78 \pm 1.34$ & $13.10_{-1.00}^{+1.00}$ & 4.610\tnote{(g)} & 14.1 \\ \midrule
EP250226a & $\checkmark$ & 337.8 & $0.83_{-0.06}^{+0.10}$ & 61.44 & $2.54_{-0.40}^{+0.43}$ & $3.21_{-0.10}^{+0.10}$ & $1.52 \pm 0.84$ & $8.68_{-3.00}^{+5.50}$ & 3.315\tnote{(h)} & 1.97\\ \midrule
\bottomrule
\end{tabular}
\tablefoot{The X-ray fluxes measured by WXT are reported in the energy band 0.5-4 keV, while the flux in $\gamma$-rays is in the energy band 10-1000 keV. For events with a measured redshift, the k-corrected isotropic-equivalent luminosity $L_{\rm iso}$ is reported in the energy range 1-$10^4$ keV.}

\tablebib{
(a)~\citet{Saccardi2024GCN, Levan2024}; (b) \citet{Jiang2025, vanHoof2026, Quirola-Vasquez2024GCN}; (c) \citet{Bochenek2024GCN}; (d) \citet{Xu2024GCN};
(e) \citet{Moskvitin2024GCN, Izzo2024GCN}; (f) \citet{Schneider2024GCN}; (g) \citet{deUgartePostigo2025GCN}; (h) \citet{Zhu2025GCN}; (0) Tentative association with a sub-threshold event in Konus-Wind (priv. comm., see Wu et al. 2026, in prep.); (1) Detected by Swift/BAT \citep{BAT_EP240315a} and Konus-Wind \citep{KW_EP240315a}; (2) Detected by Konus-Wind \citep{KW_EP240703a}; (3) Detected by Konus-Wind \citep{KW_EP240802a}, SVOM/GRM \citep{SVOM_EP240802a}, and AstroSat CZTI \citep{Astrosat_EP240802a}; (4) Detected by Konus-Wind \citep{KW_240804b}; (5) Detected by SVOM/ECLAIRs. The reported $T_{90,\gamma}$ value has been computed in the ECLAIRs energy range 4-30 keV, as the burst is only marginally detected in the 50–300 keV range. Values taken from Adrien et al., in prep.
}
\end{threeparttable}
\end{table}
\FloatBarrier

\begin{table}
\centering
\tiny
\setlength{\tabcolsep}{3pt}
\renewcommand{\arraystretch}{1.0}

\begin{threeparttable}
\caption{List of FXTs detected by Einstein Probe from February 2024 to February 2025 with a $\gamma$-ray counterpart, but whose EP observations started after the GRB (i.e. the GRB was not discovered by EP).}
\label{tab:Gamma_counterparts_butnoEPdetected}

\begin{tabular}{@{}lccccccccc@{}}
\toprule
Name  & Visible & $T_{0,\gamma}$ & $T_{90,\gamma}$ & Avg $\gamma$-ray flux & $T_{0,X\text{-}ray}$ & $T_{100,X\text{-}ray}$ & Avg X-ray flux & $z$ & WXT observation start \\
 &  by GBM? & UTC & [s] & [$10^{-7}\,\rm erg\,cm^{-2}\,s^{-1}$] & UTC & [s] & [$10^{-9}\,\rm erg\,cm^{-2}\,s^{-1}$] & & \\
\midrule
\href{https://gcn.nasa.gov/circulars/events/ep241104a}{EP241104a}
& $\checkmark$ 
& 2024-11-04T18:30:15
& 10.47
& $3.24_{-0.76}^{+0.76}$
& 2024-11-04T18:33:01
& 2555
& $0.08_{-0.02}^{+0.01}$
& -
& $\sim$3 min after $T_{0,\gamma}$ \\
\midrule

\href{https://gcn.nasa.gov/circulars/events/ep241115a}{EP241115a}
& $\times$
& 2024-11-15T05:45:36
& 103
& $1.50_{-0.21}^{+0.21}$
& 2024-11-15T05:47:22
& 1457
& $0.11_{-0.03}^{+0.03}$
& -
& $\sim$2 min after $T_{0,\gamma}$ \\
\midrule

\href{https://gcn.nasa.gov/circulars/events/ep250205a}{EP250205a}
& $\checkmark$ 
& 2025-02-05T21:25:33
& 105.47
& $0.60_{-0.02}^{+0.02}$
& 2025-02-05T21:30:14
& 907
& $0.10_{-0.01}^{+0.01}$
& 3.55\tnote{a}
& $\sim$5 min after $T_{0,\gamma}$ \\
\bottomrule

\end{tabular}

\tablefoot{Since no X-ray flux simultaneous to the $\gamma$-ray prompt emission is available, these events have been excluded from the simultaneous flux comparison.}
\tablebib{(a)~\citet{deUgartePostigo2025GCN}}

\end{threeparttable}
\end{table}
\FloatBarrier

\begingroup
\small
\setlength{\tabcolsep}{3pt}
\renewcommand{\arraystretch}{1.0}
\FloatBarrier
\clearpage
\begin{ThreePartTable}

\begin{longtable}{@{}lccccccc@{}}
\caption{List of FXTs detected by Einstein Probe from February 2024 to February 2025, without a $\gamma$-ray counterpart.}\label{tab:NoGamma_counterparts}\\
\toprule
Name & Visible by GBM? & Avg flux X-ray  & Avg $\gamma$-ray UL & Peak flux X-ray  & Peak $\gamma$-ray UL & z & $L_{\rm iso}$\\
 & & [$10^{-9} \, \rm erg/cm^{2}/s$] & [$10^{-8} \, \rm erg/cm^{2}/s$] & [$10^{-8} \, \rm erg/cm^{2}/s$] & [$10^{-7} \, \rm erg/cm^{2}/s$] & & [$10^{49} \, \rm erg/s$]\\
 & & & soft/normal/hard & & soft/normal/hard & & \\
\toprule
\endfirsthead
\toprule
Name & Visible by GBM? & Avg flux WXT  & Avg GBM UL & Peak flux WXT  & Peak GBM UL & z & $L_{\rm iso}$ \\
\midrule
\endhead
\midrule
\multicolumn{8}{r}{Continued on next page}\\
\endfoot
\bottomrule
\endlastfoot

EP240316a & $\checkmark$ & $1.23_{-0.21}^{+0.58}$ & <$0.91$/$1.25$/$2.01$ & $1.07 \pm 0.78$ & <$0.64$/$1.18$/$2.04$ & - & -\\ \midrule
EP240331a & $\times$ & $2.91_{-0.36}^{+0.99}$ & <$5.5$\tnote{1} & $1.97 \pm 1.21$ & - & - & -\\ \midrule
EP240413a & $\checkmark$ & $0.75_{-0.05}^{+0.13}$ & <$0.81$/$1.39$/$2.79$ & $1.88 \pm 1.01$ & <$0.68$/$0.90$/$1.66$ & - & -\\ \midrule
EP240414a & $\times$ & $0.42_{-0.06}^{+0.12}$ & <$17$\tnote{1} & $0.78 \pm 0.57$ & - & 0.400\tnote{a} & <$13$\\ \midrule
EP240416a & $\times$ & $0.64_{-0.09}^{+0.43}$ & <$5.1$\tnote{1} & $0.34 \pm 0.25$ & - & - & -\\ \midrule
EP240417a & $\checkmark$ & $0.11_{-0.02}^{+0.02}$ & <$0.51$/$1.03$/$1.27$ & $0.04 \pm 0.03$ & <$0.17$/$0.29$/$0.54$ & - & -\\ \midrule
EP240420a & $\times$ & $3.91_{-0.37}^{+0.60}$ & <$12$\tnote{1} & $2.19 \pm 1.07$ & - & 0.820\tnote{b} & <$50$\\ \midrule
EP240426b & $\checkmark$ & $0.44_{-0.06}^{+0.31}$ & <$1.10$/$1.71$/$2.72$ & $0.18 \pm 0.11$ & <$0.33$/$0.48$/$0.69$ & - & - \\ \midrule
EP240506a & $\checkmark$ & $2.59_{-0.37}^{+1.89}$ & <$1.44$/$2.23$/$3.94$ & $2.35 \pm 1.45$ & <$0.75$/$0.94$/$0.85$ & 0.120\tnote{c} & <$0.12$\\ \midrule
EP240618a & $\checkmark$ & $3.35_{-0.39}^{+1.27}$ & <$0.89$/$1.26$/$2.24$ & $2.16 \pm 1.30$ & <$1.07$/$1.66$/$3.21$ & - & -\\ \midrule
EP240625a & $\checkmark$ & $0.32_{-0.03}^{+0.20}$ & <$1.28$/$2.14$/$3.83$ & $0.36 \pm 0.26$ & <$1.09$/$1.74$/$2.62$ & - & -\\ \midrule
EP240626a & $\checkmark$ & $3.14_{-0.84}^{+2.78}$ & <$1.31$/$1.88$/$2.99$ & $3.16 \pm 1.94$ & <$0.82$/$1.17$/$1.19$ & - & -\\ \midrule
EP240702a & $\checkmark$ & $1.78_{-0.17}^{+0.28}$ & <$0.73$/$1.36$/$2.69$ & $3.11 \pm 1.33$ & <$0.98$/$1.47$/$2.45$ & - & -\\ \midrule
EP240708a & $\times$ & $0.08_{-0.02}^{+0.01}$ & <$10$\tnote{1} & $0.37 \pm 0.23$ & - & 0.370\tnote{d} & <$6.6$\\ \midrule
EP240806a & $\checkmark$ & $0.99_{-0.08}^{+0.24}$ & <$0.82$/$1.53$/$3.04$ & $1.13 \pm 0.69$ & <$0.79$/$1.15$/$2.00$ & 2.818\tnote{e} & <$126$\\ \midrule
EP240816a & $\checkmark$ & $0.43_{-0.06}^{+0.16}$ & <$2.29$/$2.42$/$3.16$ & $0.80 \pm 0.58$ & <$1.91$/$1.73$/$2.21$ & - & -\\ \midrule
EP240816b & $\checkmark$ & $2.85_{-0.85}^{+5.26}$ & <$1.03$/$1.56$/$3.10$ & $1.64 \pm 1.20$ & <$1.10$/$1.84$/$3.69$ & - & -\\ \midrule
EP240820a & $\times$ & $0.48_{-0.13}^{+0.12}$ & <$5.8$\tnote{1} & $1.18 \pm 0.75$ & - & - & -\\ \midrule
EP240908a & $\times$ & $0.33_{-0.02}^{+0.04}$ & <$2.9$\tnote{1} & $0.82 \pm 0.51$ & - & - & -\\ \midrule
EP240918a & $\times$ & $0.12_{-0.01}^{+0.01}$ & <$3.5$\tnote{1} & $0.45 \pm 0.28$ & - & - & -\\ \midrule
EP240918b & $\checkmark$ & $0.46_{-0.11}^{+0.10}$ & <$1.17$/$1.74$/$3.37$ & $0.20 \pm 0.12$ & <$0.27$/$0.31$/$0.73$ & - & -\\ \midrule
EP240918c & $\checkmark$ & $4.95_{-1.44}^{+5.86}$ & <$1.13$/$1.52$/$2.78$ & $8.67 \pm 4.88$ & <$0.91$/$1.33$/$2.16$ & - & -\\ \midrule
EP241021a & $\checkmark$ & $0.50_{-0.11}^{+0.13}$ & <$0.95$/$1.37$/$2.36$ & $0.14 \pm 0.07$ & <$0.23$/$0.29$/$0.47$ & 0.75\tnote{f} & <$4.6$\\ \midrule
EP241026b & $\checkmark$ & $0.19_{-0.03}^{+0.03}$ & <$0.45$/$1.14$/$2.23$ & $0.89 \pm 0.65$ & <$0.58$/$0.86$/$1.60$ & 2.294\tnote{g} & <$57$\\ \midrule 
EP241101a & $\checkmark$ & $0.97_{-0.16}^{+0.16}$ & <$0.66$/$0.96$/$1.76$ & $0.55 \pm 0.41$ & <$0.27$/$0.31$/$0.60$ & - & -\\ \midrule
EP241103a & $\checkmark$ & $4.89_{-0.81}^{+1.78}$ & <$0.78$/$1.12$/$1.83$ & $3.11 \pm 1.66$ & <$0.87$/$1.39$/$2.48$ & - & -\\ \midrule
EP241107a & $\checkmark$ & $0.45_{-0.02}^{+0.02}$ & <$2.51$/$3.22$/$4.40$ & $3.01 \pm 1.09$ & <$1.01$/$1.47$/$2.29$ & 0.456\tnote{h} & <$3.3$\\ \midrule
EP241111a & $\times$ & $0.29_{-0.05}^{+0.69}$ & $<7.40/5.68/6.75$\tnote{2} & $0.36 \pm 0.26$ & $<11.1/8.51/10.1$\tnote{2} & - & -\\ \midrule
EP241113a & $\checkmark$ & $1.08_{-0.18}^{+0.34}$ & <$0.28$/$0.31$/$0.96$ & $1.94 \pm 1.00$ & <$0.74$/$1.26$/$1.64$ & 1.530\tnote{i} & <$5.8$\\ \midrule
EP241119a & $\times$ & $0.37_{-0.04}^{+0.12}$ & <$7.0$\tnote{1} & $1.26 \pm 0.92$ & - & - & -\\ \midrule
EP241125a & $\checkmark$ & $1.52_{-0.44}^{+18.35}$ & <$0.79$/$1.42$/$2.98$ & $0.89 \pm 0.69$ & <$0.29$/$0.54$/$1.05$ & - & -\\ \midrule
EP241126a & $\times$ & $5.08_{-0.65}^{+1.22}$ & - & $3.69 \pm 1.77$ & - & - & - \\ \midrule
EP241201a & $\times$ & $26.33_{-17.06}^{+52.05}$ & <$1.9$\tnote{1} & $18.58 \pm 15.32$ & - & - & - \\ \midrule
EP241202b & $\checkmark$ & $0.84_{-0.15}^{+0.17}$ & <$0.71$/$1.21$/$2.49$ & $0.51 \pm 0.37$ & <$0.34$/$0.57$/$0.95$ & - & -\\ \midrule
EP241208a & $\times$ & $1.81_{-0.41}^{+7.03}$ & <$9.8$\tnote{1} & $1.30 \pm 0.96$ & - & - & - \\ \midrule
EP241217a & $\times$ & $0.25_{-0.04}^{+0.04}$ & <$3.6$\tnote{1} & $0.89 \pm 0.57$ & - & 4.590\tnote{l} & $951$\\ \midrule
EP241223a & $\checkmark$ & $0.81_{-0.09}^{+1.10}$ & <$0.99$/$1.42$/$2.28$ & $0.47 \pm 0.28$ & <$0.38$/$0.61$/$1.02$ & - & -\\ \midrule
EP241231a & $\checkmark$ & $0.43_{-0.11}^{+0.10}$ & <$0.85$/$1.39$/$2.80$ & $0.16 \pm 0.10$ & <$0.15$/$0.26$/$0.72$ & - & -\\ \midrule
EP241231b & $\checkmark$ & $0.35_{-0.07}^{+0.08}$ & <$0.67$/$1.35$/$2.86$ & $0.34 \pm 0.25$ & <$0.40$/$0.80$/$1.78$ & - & -\\ \midrule
EP250108a & $\checkmark$ & $0.13_{-0.03}^{+0.40}$ & <$0.85$/$1.83$/$5.30$ & $0.03 \pm 0.02$ & <$0.07$/$0.06$/$0.08$ & 0.176\tnote{m} & <$0.22$\\ \midrule
EP250111a & $\times$ & $1.18_{-0.15}^{+0.25}$ & - & $2.80 \pm 1.28$ & - & - & -\\ \midrule
EP250125a & $\times$ & $0.21_{-0.02}^{+0.05}$ & $<0.44/0.70/1.39$\tnote{2} & $1.71 \pm 0.87$ &
$<0.91/1.39/2.73$\tnote{2} & $2.890$\tnote{n} &
$<61$\\ \midrule
EP250207b & $\times$ & $0.46_{-0.12}^{+0.12}$ & - & $0.48 \pm 0.36$ & - & 3.200\tnote{o} & -\\ \midrule
EP250208b & $\checkmark$ & $0.15_{-0.03}^{+0.03}$ & <$1.42$/$1.54$/$2.72$ & $0.10 \pm 0.06$ & <$0.22$/$0.34$/$0.57$ & - & -\\ \midrule
EP250212a & $\checkmark$ & $1.01_{-0.11}^{+0.25}$ & <$0.72$/$1.16$/$1.90$ & $1.33 \pm 0.97$ & <$0.83$/$1.37$/$2.33$ & - & -\\ \midrule
EP250220a & $\times$ & $0.42_{-0.12}^{+0.12}$ & - & $0.43 \pm 0.33$ & - & - & -\\ \midrule
EP250223a & $\checkmark$ & $0.20_{-0.02}^{+0.07}$ & <$0.97$/$1.58$/$2.85$ & $1.79 \pm 0.81$ & <$0.65$/$1.28$/$2.56$ & 2.756\tnote{p} & <$123$\\ \midrule
EP250225a & $\times$ & $0.18_{-0.05}^{+0.04}$ & - & $0.28 \pm 0.21$ & - & - & -\\ \midrule
EP250227a & $\times$ & $1.52_{-0.25}^{+1.79}$ & - & $0.90 \pm 0.55$ & - & - & -\\ \midrule

\end{longtable}
\tablefoot{The X-ray fluxes are measured by WXT and reported in the energy band 0.5-4 keV, the flux upper limits in $\gamma$-rays are measured by GBM (unless otherwise stated) and are reported in the energy band 10-1000 keV, for three spectral templates (soft, normal and hard, as described in Sect. \ref{sec:methodsGBM}). Events whose upper limit comes from \citealt{Zhang2025_upperlimits} or the \textit{SVOM}/GRM targeted search are indicated with the superscript (1) or (2). For events with a measured redshift, the upper limit on the k-corrected isotropic-equivalent luminosity $L_{\rm iso}$, computed from the flux upper limit of the \textit{normal} template, is reported in the energy range 1-$10^4$ keV.}
\tablebib{(1) From \citet{Zhang2025_upperlimits}; (2) From \textit{SVOM}/GRM targeted search; (a) \citet{Jonker2024GCN, vanDalen2025}; (b) \citet{An2024GCN};
(c) \citet{Liang2026}; (d) \citet{vanHoof2026b}; (e) \citet{Quirola2024GCN}; (f) \citet{Pugliese2024GCN};
(g) \citet{Joyce_26b}; (h) \citet{Adami2024GCN_241107a,Quirola2024GCN_241107a}; (i) \citet{Quirola2024GCN_241113a}; (l) \citet{Levan2024GCN_241217a}; (m) \citet{Zhu2025GCN_250108a}; (n) \citet{Levan2025GCN_EP250125a}; (o) van Hoof et al. 2026, in prep.; (p) \citet{Levan2025GCN_EP250223a}.}
\end{ThreePartTable}
\endgroup

\begin{table}[H]
    \centering
    \caption{List of FXTs detected by Einstein Probe from February 2024 to February 2025, visible by LAT.}\label{Table_LAT}
    \begin{tabular}{llllllll}
    \toprule
        Source Name & Peak/Avg. & T100 [s] & LAT-WXT & N$_{ph}$ & WXT Peak & LAT avg UL & LAT peak UL \\ 
                    &        WXT Flux       &          &                Coverage [\%]       &          & Covered                 &    [$10^{-9}$ erg/cm2/s] (TS)   & [$10^{-6}$ erg/cm2/s] (TS)   \\ \hline\hline
        EP240219a & 34.1 & 449.95 & 80.5 & 18 & $\times$ & 2.75 & -   \\ \hline
        EP240413a & 38.4 & 375.35 & 100 & 6 & \checkmark & 1.87 & 1.1  \\ \hline
        EP240618a & 8.4 & 69.85 & 100 & 1 & \checkmark & 10.2 & 1.34  \\ \hline
        EP240801a & 29.1 & 266.35 & 100 & 4 & \checkmark & 3.24 & 1.12  \\ \hline
        EP240816b & 10.5 & 52.15 & 100 & 0 & \checkmark & 8.83 & 0.92  \\ \hline
        EP241025a & 82.3 & 1939.6 & 18.6  & 15 & \checkmark & 4.59 (7) & 1.05  \\ \hline
        EP241107a & 299.7 & 1275.95 & 83.8 & 14 & \checkmark & 0.80 & 0.74  \\ \hline
        EP241113a & 30.9 & 241.9 & 63.5 & 3 & $\times$ & 4.04 & - \\ \hline
        EP241231a & 7.4 & 129.55 & 100 & 3 &  \checkmark & 2.94 & 0.09  \\ \hline
        EP250108a & 6.5 & 1428.15 & 53.2 & 3 & \checkmark & 0.69 & 0.02  \\ \hline
        EP250215a & 641.2 & 1673.05 & 48.7 & 17 & \checkmark & 1.01 & 0.76  \\ \hline
        EP250223a & 397.8 & 1205.5 & 72.1 & 27 & \checkmark & 0.72 & 1.42 \\ \hline
    \end{tabular}
    \tablefoot{The sources in the list satisfy one of the two conditions: 1. LAT-WXT coverage > 50\%, or 2. LAT covered the WXT-peak time. These LAT upper limits are estimated at a fixed spectral index of -2 (Normal). The ULs derived using the hard (soft) spectral template are, on average, approximately 1.1 (0.9) times the ones obtained with the normal spectral index.}
\end{table}

\twocolumn

\section{Additional plots}

\begin{figure}[H]
   \centering
   \includegraphics[width=0.49\textwidth]{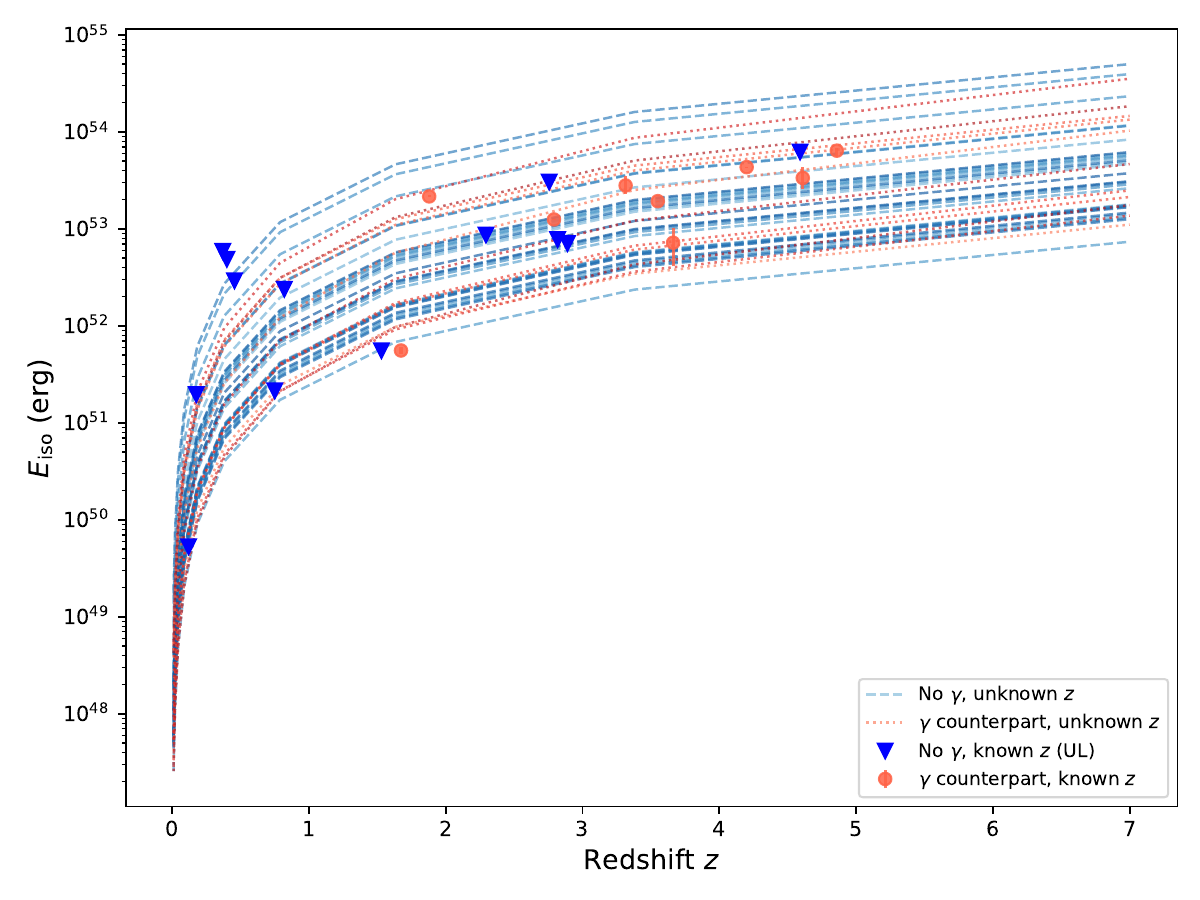}
   \includegraphics[width=0.49\textwidth]{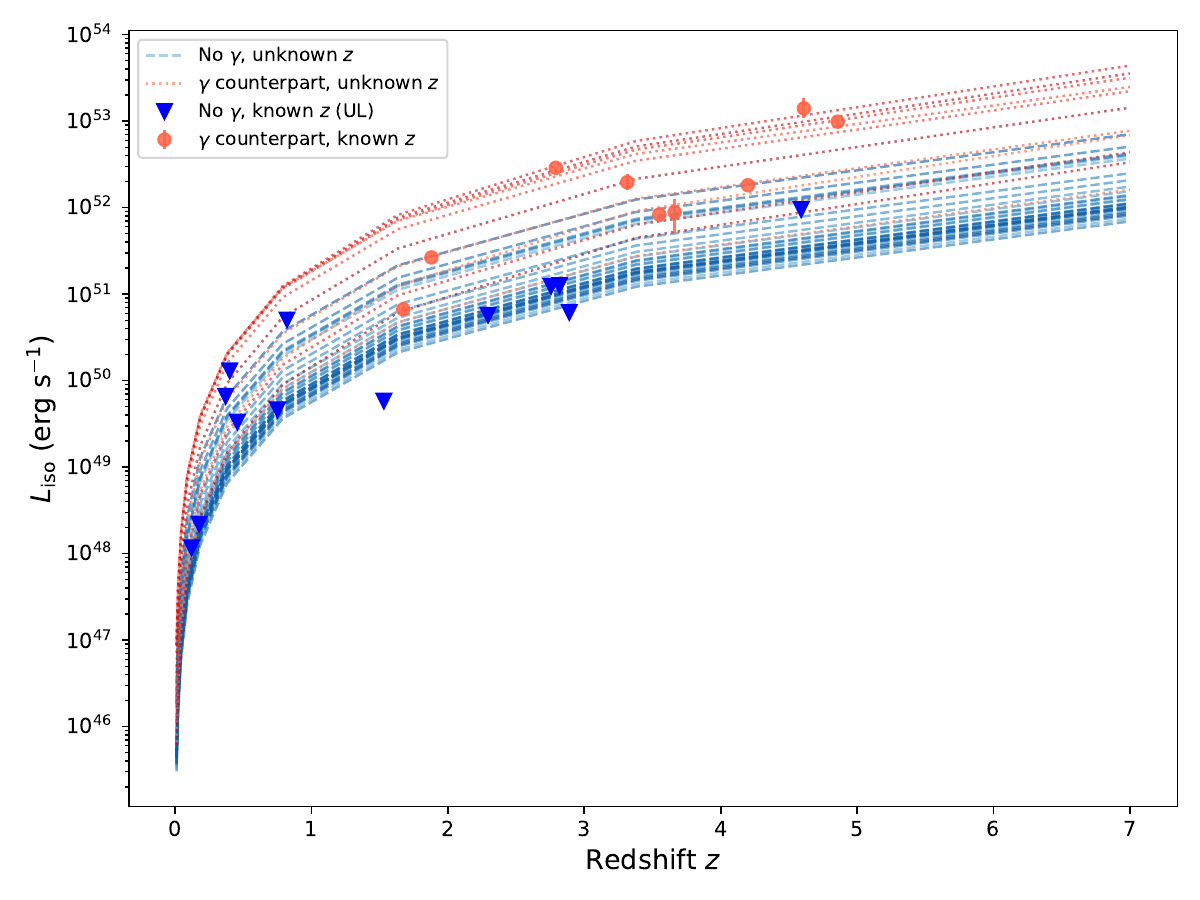}
      \caption{Isotropic equivalent energy $E_{\rm iso}$ (\textit{top panel}) and isotropic equivalent luminosity $L_{\rm iso}$ (\textit{bottom panel}) as a function of redshift for the EP transients analysed in this work. Red points indicate EP sources with a confirmed $\gamma$-ray counterpart and known redshift, for which $L_{\rm iso}$ and $E_{\rm iso}$ are computed from the measured prompt flux in the 10--1000\,keV band ($k$-corrected to the 1--$10^4$\,keV rest-frame) over the $\gamma$-ray duration $T_{90,\gamma}$. Blue downward arrows indicate $3\sigma$ upper limits for EP sources without a detected $\gamma$-ray counterpart and known redshift, derived using the normal spectral template; for these events, $E_{\rm iso}$ is computed using the X-ray duration $T_{\rm 100,X\text{-}ray}$. Blue and red curves show the values of $L_{\rm iso}$ and $E_{\rm iso}$ corresponding to the flux measurements and upper limits of EP sources without and with a $\gamma$-ray counterpart, respectively, for events lacking a redshift measurement. Sources with redshift, but no coverage in $\gamma$-rays, are not represented in this plot.}
         \label{fig:EisoLiso_vs_z}
\end{figure}

\begin{figure}[H]
   \centering
   \includegraphics[width=0.5\textwidth]{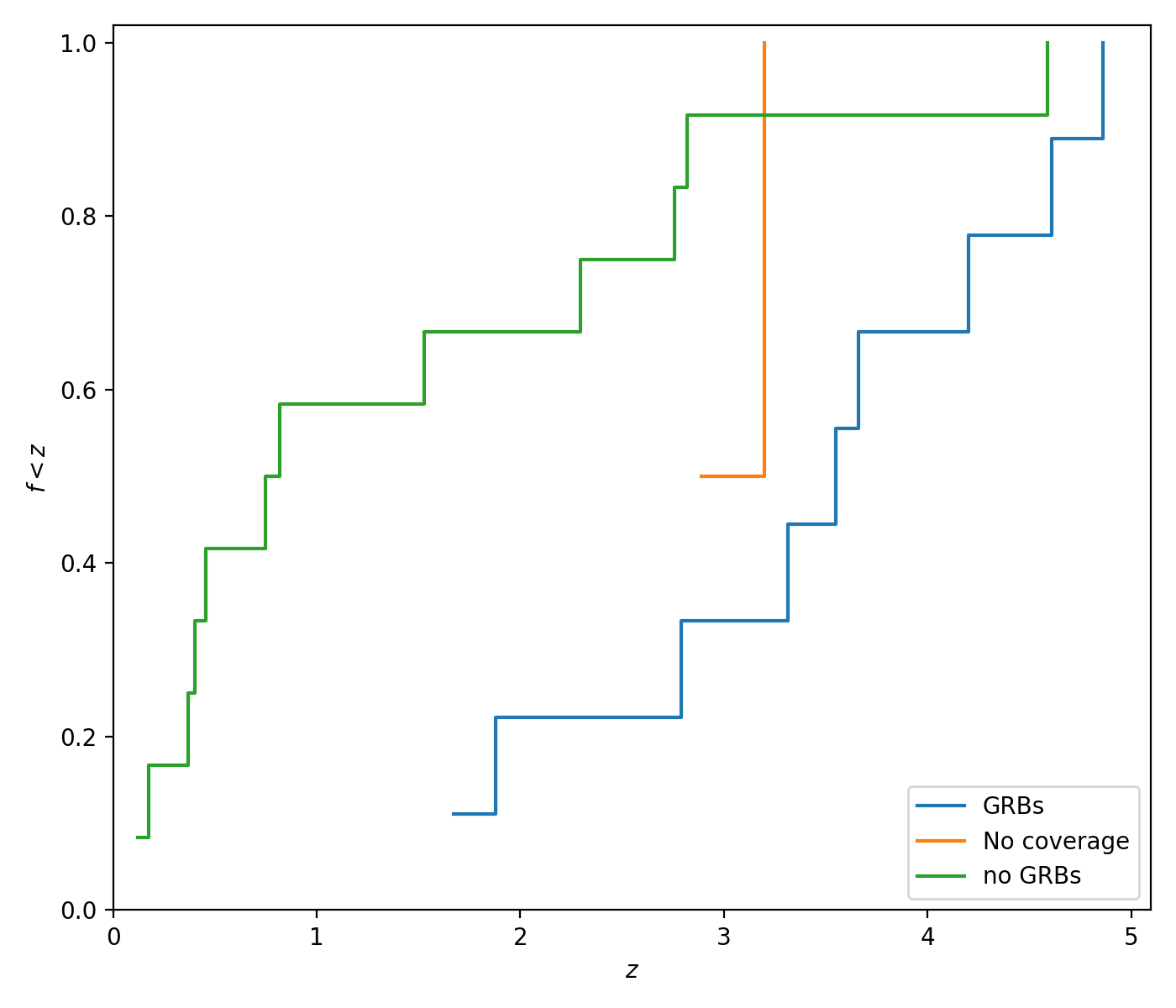}
      \caption{Cumulative redshift distribution of the first-year EP sample, divided into events with an associated GRB (blue), events without a $\gamma$-ray counterpart despite gamma-ray coverage (green, 'no GRBs'), and events lacking any gamma-ray coverage (orange, 'No coverage'). 
      }
         \label{fig:cumulative_z}
\end{figure}

\begin{figure}[H]
   \centering
   \includegraphics[width=0.5\textwidth]{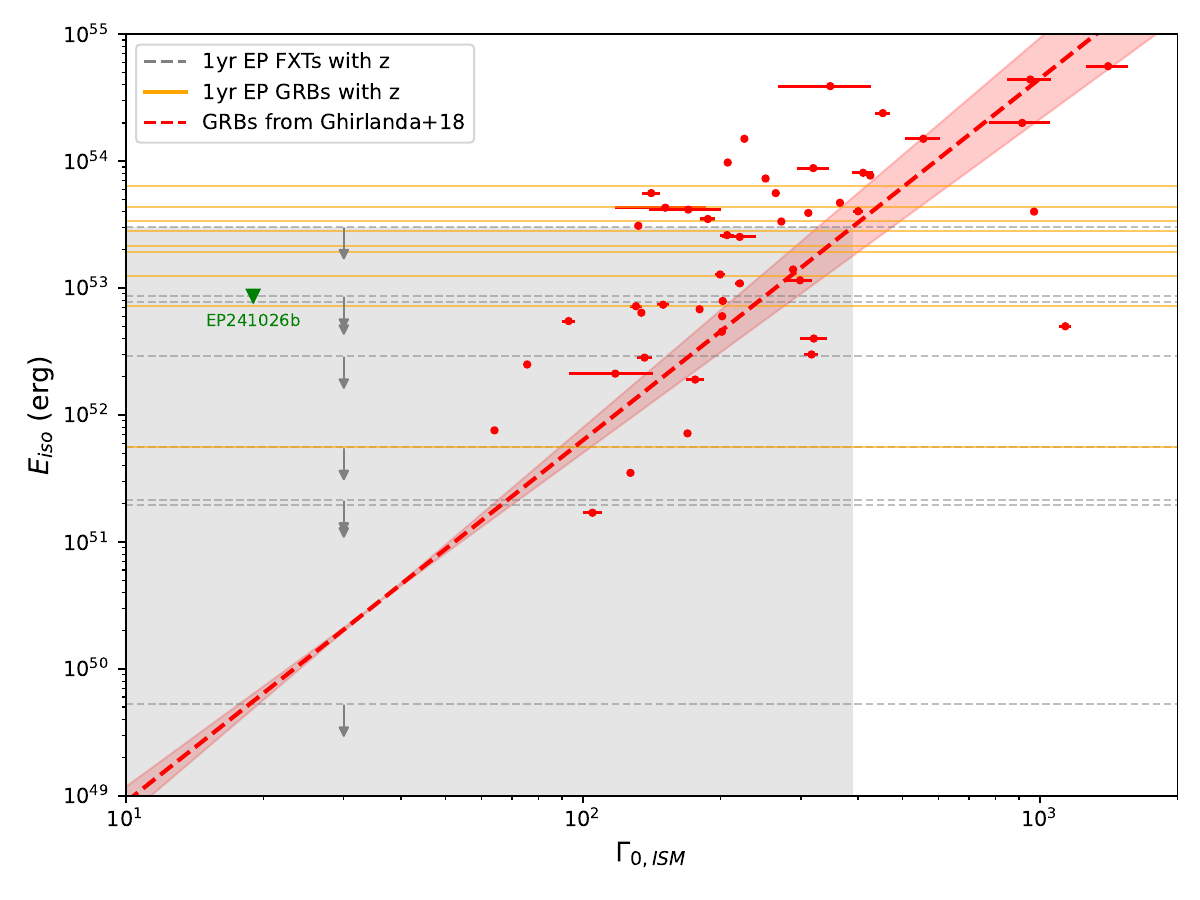}
      \caption{Correlation between the bulk Lorentz factor $\Gamma_0$ and the isotropic equivalent energy $E_{\rm iso}$, as derived in \citep{Ghirlanda2018}. For the EP sources with and without $\gamma$-ray counterpart, orange solid lines indicate $E_{\rm iso}$ measurements, while grey dashed lines represent upper limits, respectively.}
         \label{fig:GG18_correlations_appendix}
\end{figure}

\end{appendix}

\end{document}